\documentclass[useAMS,usenatbib]{lib/mnras}
\usepackage{natbib}
\usepackage{epsfig}
\usepackage{scrtime}
\usepackage{graphicx}	% Including figure files
\usepackage{amsmath}	% Advanced maths commands
\usepackage{amssymb}	% Extra maths symbols
\usepackage{float}
\usepackage[caption = false]{subfig}
\usepackage{multirow}
\usepackage{rotating}
\usepackage{ulem}
\usepackage{booktabs}
\usepackage{caption}

\newcommand{\Oiii}{\mathrm{O\ III}}

\newcommand{\lsigG}{$L - \sigma$}

\newcommand{\Msol}{M$_\odot$}
\newcommand{\mincir}{\raise-3.truept\hbox{\rlap{\hbox{$\sim$}}\raise4.truept\hbox{$<$}\ }}

\title[Independent cosmological constraints from HIIG. KMOS data]{Independent cosmological constraints from high-z HII~galaxies: new results from VLT-KMOS data}
\author[Ana Luisa Gonz\'alez-Mor\'an et al.] {Ana Luisa Gonz\'alez-Mor\'an$^{1}$\thanks{Contact e-mail: analuisagm@inaoep.mx},
Ricardo Ch\'avez$^{2}$\thanks{Contact e-mail: r.chavez@irya.unam.mx},
Elena Terlevich$^{1}$, \newauthor
Roberto Terlevich$^{1,3}$, 
David Fern\'andez-Arenas$^{4}$,
Fabio Bresolin$^{5}$,
Manolis Plionis$^{6,7}$, \newauthor
Jorge Melnick$^{8,9}$,
Spyros Basilakos$^{10}$ and
Eduardo Telles$^{9}$.
\\ \\
$^{1}$ Instituto Nacional de Astrof\'\i sica, \'Optica y Electr\'onica,Tonantzintla, C.P. 72840, Puebla, M\'exico \\
$^{2}$ CONACYT-Instituto de Radioastronom\'ia y Astrof\'isica, UNAM, Campus Morelia, C.P. 58089, Morelia, M\'exico\\
$^{3}$ Institute of Astronomy, University of Cambridge, Cambridge, CB3 0HA, UK\\
$^{4}$ Kavli Institute for Astronomy and Astrophysics, Peking University, Beijing 100871, China \\
$^{5}$ Institute for Astronomy, University of Hawaii, 2680 Woodlawn Drive, 96822 Honolulu,HI USA \\
$^{6}$ National Observatory of Athens, P. Pendeli, Athens, Greece \\
$^{7}$ Physics Dept., Aristotle Univ. of Thessaloniki, Thessaloniki 54124, Greece \\
$^{8}$ European Southern Observatory, Santiago de Chile, Chile \\
$^{9}$ Observatorio Nacional, Rua Jos\'e Cristino 77, 20921-400 Rio de Janeiro, Brasil\\
$^{10}$ Academy of Athens Research Center for Astronomy \& Applied Mathematics, Soranou Efessiou 4, 11-527 Athens, Greece \\
}

\begin{document}

\date{v13 --- Compiled at \thistime\ hrs  on \today\ }

\pagerange{\pageref{firstpage}--\pageref{lastpage}} \pubyear{2021}

\maketitle

\label{firstpage}

\begin{abstract}

We present independent determinations of cosmological parameters using the distance estimator based on the established correlation between the Balmer line luminosity, L(H$\beta$), and the velocity dispersion ($\sigma$) for HII galaxies (HIIG). These results are based on new VLT-KMOS  high spectral resolution observations of 41 high-z ($1.3 \leq$ z $\leq 2.6$) HIIG combined with published data for 45 high-z and 107 z $\leq 0.15$ HIIG, while the cosmological analysis is based on the MultiNest MCMC procedure not considering systematic uncertainties. Using only HIIG to constrain the matter density parameter ($\Omega_m$), we find $\Omega_m = 0.244^{+0.040}_{-0.049}$ (stat), an improvement over our best previous cosmological parameter constraints, as indicated by a 37\% increase of the FoM. The marginalised best-fit parameter values for the plane $\{\Omega_m; w_0\}$ = $\{0.249^{+0.11}_{-0.065}; -1.18^{+0.45}_{-0.41}\}$ (stat) show an improvement  of the cosmological parameters constraints by 40\%. Combining the HIIG Hubble diagram, the cosmic microwave background (CMB) and the baryon acoustic oscillation (BAO) probes yields $\Omega_m=0.298 \pm 0.012$ and $w_0=-1.005 \pm 0.051$, which are certainly compatible --although less constraining-- than the solution based on the joint analysis of SNIa/CMB/BAO. An attempt to constrain the evolution of the dark energy with time (CPL model), using a joint analysis of the HIIG, CMB and BAO measurements, shows a degenerate 1$\sigma$ contour of the parameters in the $\{w_0,w_a\}$ plane.

\end{abstract}

\begin{keywords}
galaxies: starburst -- cosmology: dark energy -- cosmological parameters -- observations.
\end{keywords}

\section{Introduction}

An accelerated cosmic expansion was observed two decades ago suggesting the presence of a non-zero cosmological constant $\Lambda$ \citep{Riess1998, Perlmutter1999}. The large vacuum energy density required is usually referred to as  dark energy (DE).

High redshift objects are interesting for cosmology not only because of their large range of distances but also because they contain important information about physical processes in the early Universe while providing constraints on the components of the present Universe. This is particularly so when the joint analysis at z $\lesssim 2$ for SNIa (\citealt{Riess1998,Perlmutter1999,Hicken2009,Amanullah2010,Riess2011,Suzuki2012,Betoule2014,Scolnic2018}), BAO \citep[e.g.][]{Jaffe2001, Pryke2002, Spergel2007} and independent cosmological parameters determination by means of HII galaxies (HIIG) observations \citep[e.g.][]{Chavez2012, Terlevich2015, Chavez2016, Fernandez2018, GonzalezMoran2019} with the results at z $\sim$ 1000 from the CMB fluctuations (e.g. \citealt{PlanckCollaboration2014, PlanckCollaboration2016a}) are performed. Individual solutions in the plane $\{\Omega_m; w_0\}$ are degenerate and only by combining them one can break the degeneracy.s

Most of the mass-energy in the Universe is due to its least understood components DE and dark matter (DM). The contribution of  stars, planets and interstellar matter, the components we best understand, is almost negligible \citep[see e.g.][for a discussion and description of the methods for deriving these components of the total cosmic mass density]{FukugitaPeebles2004}.

Extensive observing programs at high redshift need to be carried out to determine with higher confidence the form of the DE Equation of State (EoS) and to decide whether the $w$ parameter (relation between the pressure $p$ and the mass-energy density $\rho c^2$ in the DE EoS) evolves with look-back time \citep{PeeblesRatra1988, Wetterich1988}. Constraining cosmological parameters and confirming the results through different and independent methods should conduce to a more precise and robust cosmological model.

HIIG are compact low mass systems (M $ < 10^9$ \Msol) with their luminosity almost completely dominated by a young (age $<$ 5 Myr) massive burst of star formation \citep{Chavez2014}. By selection they are the population of extragalactic systems with the strongest narrow emission lines ($\sigma < 90$ km/s) and represent the youngest systems that can be studied in any detail. We first called them HIIG to underline  the fact that their integrated optical spectrum is completely dominated by that of a giant HII region in spite of being hosted by a compact dwarf galaxy. This is similar to what happens with QSO data where the underlying galaxy spectrum is extremely difficult to detect.

HIIG rest-frame optical spectra are dominated by strong narrow  emission lines  superimposed on a faint blue continuum hence they are easily observed up to large distances. This makes them powerful tools for studying recent star formation at high redshift using  currently available infrared instrumentation up to \mbox{z $\sim 4$}; with incoming instruments like NIRSpec \citep{Dorner2016} on the JWST \citep{Gardner2006} it will be possible  to explore them up to z $\sim$ 6.5 using H$\alpha$ or z $\sim$ 9 with the H$\beta$ and \mbox{[$\Oiii$]$\lambda\lambda$4959,5007 \AA} doublet  group. So, we are very close to observing luminous HIIG to a look-back time  corresponding to the epoch when perhaps the first of these objects were formed. 

It has been shown that  HIIG and Giant Extragalactic HII Regions (GHIIR) satisfy a correlation between the Balmer line luminosity L(H$\beta$) and the velocity dispersion ($\sigma$) of the emission lines that can be used as a cosmological distance indicator, the \lsigG\ relation \citep{Terlevich_Melnick1981, Melnick_Terlevich_Moles1988, Bordalo_Telles2011, Chavez2014}. They can potentially be observed up to very large distances, which opens the possibility of applying the \lsigG\ distance estimator to map the Hubble flow over an extremely wide redshift range.

The use of the \lsigG\ relation as a distance indicator has already been proved \citep[e.g.][and references therein]{Melnick2000, Siegel2005, Plionis2011, Chavez2012, Chavez2014, Terlevich2015, Chavez2016, GonzalezMoran2019}. The \lsigG\ relation has been used in the local Universe to significantly constrain the value of the Hubble constant, $H_0$ \citep{Chavez2012, Fernandez2018} allowing to contribute to the $H_0$ tension discussion now at the 3.1$\sigma$ level between the results from SNIa given by \cite{Riess2016} of $H_0$ = 73.24 $\pm$ 1.74 km s$^{-1}$ Mpc$^{-1}$ and the value obtained by \cite{PlanckCollaboration2016a} of $H_0$= 67.8 $\pm$ 0.9 km s$^{-1}$ Mpc$^{-1}$. The most recent $H_0$ determination based on HIIG lies in 71.0 $\pm$ 2.8(random) $\pm$ 2.1(systematic) km s$^{-1}$ Mpc$^{-1}$ \citep{Fernandez2018}. For the early Universe, we have obtained  independent determinations of the cosmological parameters $\{\Omega_m, w_0, w_a\}$ using a sample of HIIG in a redshift range of 1.3 $<$ z $<$ 2.5 observed with MOSFIRE at the Keck telescope \citep{GonzalezMoran2019}. We found constraints that are in excellent agreement with those of similar analyses using SNIa.

\citet{Plionis2011} using extensive Monte Carlo simulations predicted that just with a few tens of HIIG at high redshift, even with a large distance modulus uncertainty, one can reduce significantly the cosmological parameters solution space. In fact, they found that a reduction ($\sim$ 20 - 40 \%) of the current level of HIIG-based distance modulus uncertainty would not provide a significant improvement in the derived cosmological constraints; it is more efficient to increase instead the number of tracers.

In this paper we use a new set of high  resolution spectrophotometric observations of high redshift HIIG obtained with KMOS at the ESO VLT to improve the constraints in the parameters space of the DE EoS and $\Omega_{m}$  on the crucial range of intermediate redshift 1.3 $<$ z $<$ 2.6.

The structure of the paper is as follows: in \S \ref{sec:KMOS Observations} we present the observations and data reduction. The data are analysed in \S \ref{sec:Analysis}. Results and systematic uncertainties are discussed in \S \ref{sec:Results}. Finally, the conclusions are given in \S \ref{sec:Conclusions}.

%%%%%%%%%%%%%%%%%% Figure 1 a y b %%%%%%%%%%%%%
\begin{figure*}
\begin{center}$
\begin{array}{cc}
  \subfloat[ UDS and GOODS-S fields] {\includegraphics[width=.5\textwidth]{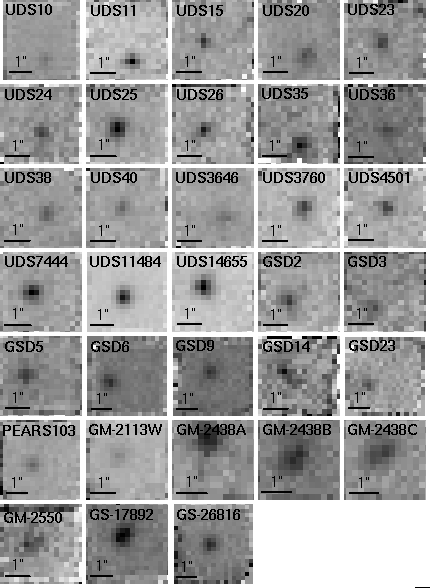}} &
  \subfloat[Q2343 and COSMOS fields] {\hspace{-0.3cm}\includegraphics[width=.5\textwidth]{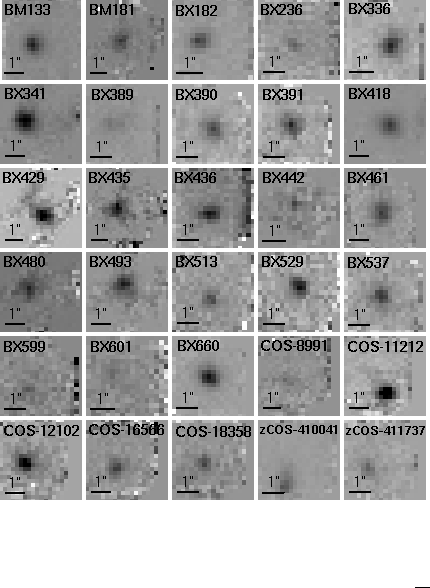}}
\end{array}$
\end{center}
	\caption{Emission line images built from KMOS IFU H band data cubes for the UDS, GOODS-S, Q2343 and COSMOS fields. The images represent the [$\Oiii$]$\lambda$5007 line for objects at z $\sim$ 2.3 and H$\alpha$ for those at z $\sim$ 1.5. The number of wavelength slices used to construct the images depends on the intensity and width of the observed emission. The KMOS data cube for GMASS-2438 presents emission at three different positions inside the IFU FoV.}
\label{KMOS sample images}
\end{figure*}

\section{KMOS observations}
\label{sec:KMOS Observations}

Large databases, mostly cosmological fields, containing HIIG at high redshifts already exist in the literature so we have the possibility of selecting appropriate candidates with the specific requirements that we need for follow-up observations. The high number density of HIIG at high redshift allows the use of multi-object spectrographs reducing notably the amount of time required to build up a significant sample. 
The last column in Table \ref{tab:table1}, where we list the objects observed, gives the source for the selected candidates.

High spectral resolution near-IR spectra  of 96 HIIG candidates were obtained using 55 hours of total observing time (PI R. Terlevich. Programmes 097.A-0039 and 098.A-0323) with the K-band Multi Object Spectrograph \citep[KMOS;][]{Sharples2013}, a second-generation instrument at the Nasmyth focal plane of the VLT at ESO, which is able to perform  simultaneous  near-infrared Integral Field Spectroscopy for 24 targets.

\subsection{Data description}
\label{sec:Data description}

The sample of 96 star-forming galaxies was selected from \citet{Erb2006a, Erb2006b, ForsterSchreiber2009, Mancini2011, van_der_Wel2011, Xia2012, Maseda2013, Maseda2014} following the criteria described in detail in \citet{GonzalezMoran2019}. The candidates have  high rest-frame equivalent width (EW) in their emission lines in a range of $1.2 <$ z $< 1.7$ and $1.9 <$ z $< 2.6$ in order to observe either H$\alpha$ or H$\beta$ and \mbox{[$\Oiii$]$\lambda$5007\AA} lines in the H band and with the objects being in dense enough fields so that at least 10 of them would fit in the field of view (FoV) of the spectrograph.

KMOS is equipped with 24 integral field units (IFUs) that can be deployed by robotic arms to positions within the patrol field of 7.2\arcmin\ in diameter. Each IFU has a square FoV of 2.8\arcsec\ x 2.8\arcsec\ sampled spatially at 0.2\arcsec\ whilst maintaining Nyquist sampling ($\sim$ 2 pixel) of the spectral resolution element at the detector.

It provides a wavelength coverage of 1.456 - 1.846 $\mu$m in the H band at the  FoV centre and achieves a spectral resolution of R = 4,000 in the atmospheric band. Due to  spectral curvature, the IFUs at the edges of the array have a slightly different wavelength coverage than those  at the centre.

We allocated the pick-off arms to specific targets in the patrol field using the KMOS Arm Allocator \citep[KARMA;][]{Wegner2008}, an automated tool that optimises the assignments taking into account  target priorities and mechanical constraints for the arms reach.

The data were obtained in service mode from June 2016 to July 2017 for period 97A  with a total of 16 Observing Blocks (OBs) distributed in 2 fields on the cosmological field Q2343 \citep{Steidel2004,Erb2006a,Erb2006b} and from December 2016 to October 2017 for the 98A period, distributed in 5 FoV  on 3 cosmological fields: the Ultra Deep Survey \citep[UDS;][]{Lawrence2007, Cirasuolo2007}, GOODS-South Deep \citep[GSD;][]{Giavalisco2004} and the Cosmic Evolution Survey \citep[COSMOS;][]{Scoville2007, Koekemoer2007}, each OB with $\sim$ 40 minutes exposure time.

The observing mode adopted was `nod to sky'. Here, the 24 targets are observed in every pair of pointings with most of the arms on targets at the first position and the remainder on targets during the second exposure after a nod. In this case, the sky background is removed simply by subtracting alternate exposures for each arm. 
The data for each IFU are processed independently.

The observed sample is presented in Table \ref{tab:table1}. The target name is given in the first column, the coordinates in the second and third columns, the cosmological field that each object belongs to in column 4, the seeing in arcseconds corrected by airmass during the field observations in column 5, the total exposure time per target in seconds in column 6, and the reference from where the candidates were selected in column 7.

\subsection{Data Reduction}
\label{sec:Data Reduction}

The sample from the Q2343 field was observed in two overlapping FoV, 16 OBs for 8 repeated targets and 8 OBs for the rest. Five FoV were observed during the 98A period.  Two FoV with 8 OBs each for the UDS field, two FoV with 8 OBs and 7 OBs for the GOODS-S field and one FoV with 8 OBs for the COSMOS field. In total 55 OBs with different position angles and  5 exposures each  were observed. The total exposure time per target is shown in Table \ref{tab:table1}.

The data reduction was carried out via the KMOS Reflex workflow\footnote{http://www.eso.org/sci/software/pipelines/reflex workflows} software \citep{Freudling2013} developed by ESO and the instrument consortia \citep{Davies2013}. 
The software corrects the frames for their dark level and structure, flat-field, computes a wavelength solution, applies an illumination correction, a standard star flux calibration and telluric correction, and finally creates a cube reconstruction of the science data.

In order to guarantee  the same orientation in all exposures for each FoV combination, we used  `kmo rotate' inside the ESO Recipe Execution tool, EsoRex\footnote{http://www.eso.org/sci/software/cpl/esorex.html} for each exposure of the OBs per FoV.

By design, the KMOS workflow combines only the 5 exposures associated to a single OB, so we reduced all the OBs separately and then  executed  `kmo combine' inside EsoRex.

The 1D spectrum extraction was made using the fits file viewer QFitsView\footnote{https://www.mpe.mpg.de/$\sim$ott/QFitsView/} \citep{Ott2012} developed at the Max Planck Institute for Extraterrestrial Physics (MPE) and included in ESO's SciSoft releases. This software allows to see in real time integrated spectra from different groups of spaxels of the KMOS data cube on  the displayed image of the spatial FoV.

%%%%%%%%%%%%%%%%%%% Table %%%%%%%%%%%%%%
\begin{table}
\caption{Description of the analysed sample.}
\label{tab:Description}
\centering
\begin{tabular}{ l l l }
\hline \hline
Sample & Description & N \\
\hline
S1   &  Observed KMOS sample &  96 \\
S2   & S1 with emission lines detection  &  61 \\
S3   & S2 with enough S/N       &        54 \\
S4   & S3 with $\log \sigma - \epsilon_{\log \sigma} \leq$1.83  &  41 \\
S5   & S4 joint with linking data (see \S \ref{Repeated observations})  & 29 \\
\hline
\hline
\end{tabular}
\end{table}

%%%%%%%%%%%%%%%%%% Figure  2 %%%%%%%%%%%%%%%%%%
\begin{figure}
\hspace*{0.4cm}
	\includegraphics[width=0.93 \columnwidth]{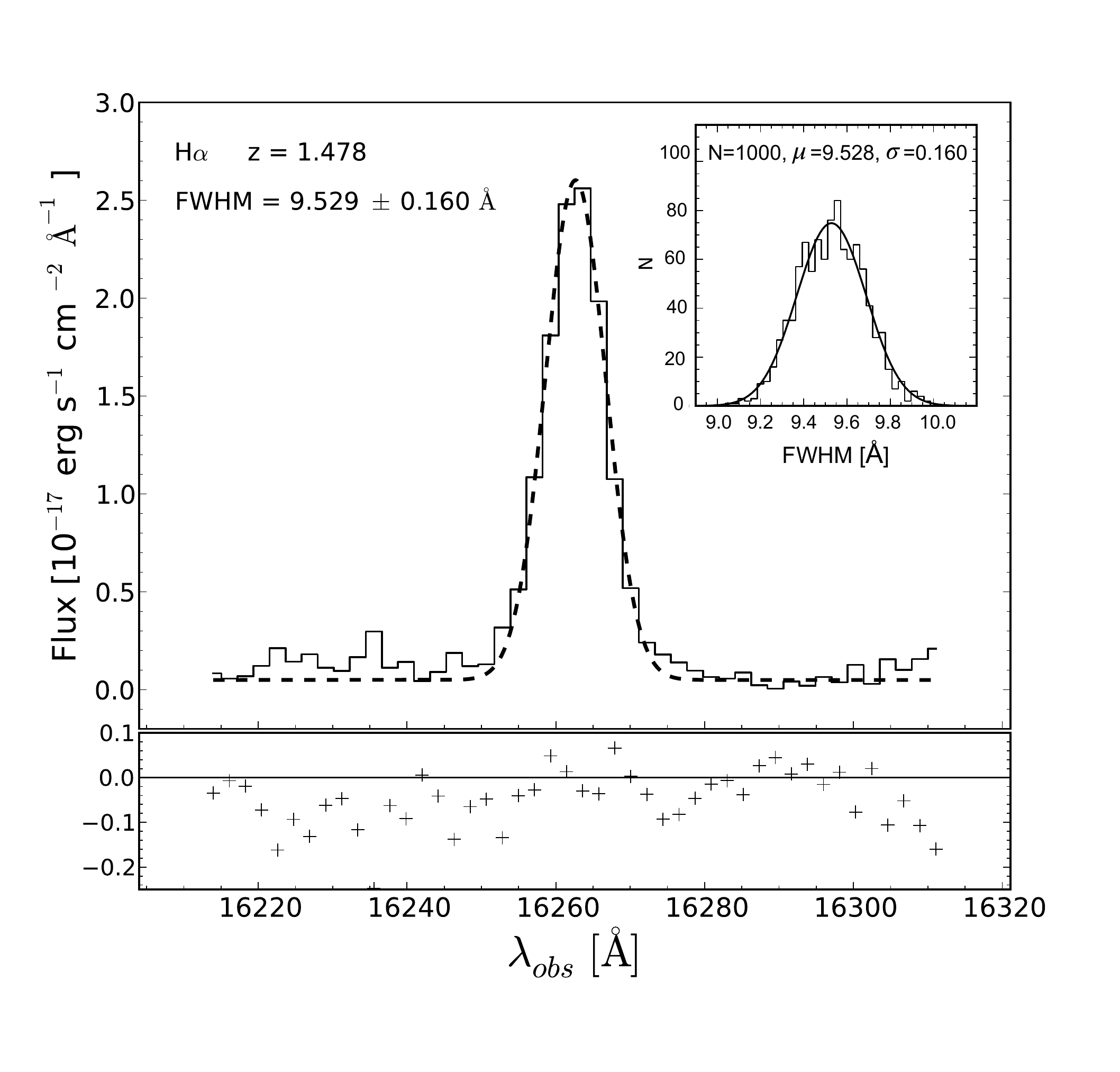}
\vspace*{-0.55cm}	
    \caption{H$\alpha$ line for Q2343-BM133. The  grey line is the spectrum, the dashed dark line is the Gaussian fit to the emission line and the box underneath shows the residuals. The inset at the upper right corner is the Monte Carlo analysis performed to the line where the standard deviation of the resulting distribution is taken as the uncertainty of the measured FWHM of the Gaussian fit.}
    \label{fig:montecarlo analysis}
\end{figure}

\section{Analysis}
\label{sec:Analysis}

Emission lines were detected in 61 (sample S2; see Table \ref{tab:Description}) of the 96  HIIG candidates (sample S1) observed. Emission line images of the S2 objects are shown in Fig. \ref{KMOS sample images}, where there are 63 images instead of 61 because the data cube for GMASS-2438 presents emission at three different positions inside the IFU FoV. The images were built from the combination of slices within the particular wavelength range where the emission line was detected in the data cube. Note that most of them have sub arcsecond diameters and appear to be single.

One possible reason for the non detections is that the redshift range of the candidates, selected from their photometric redshift and with high [$\Oiii$] equivalent width, is wider than the KMOS line detection window. We adopted the strategy that if no emission lines were detected in a single exposure, then even if emission was detected in the complete exposure time the S/N would not be sufficient to determine the line width with the needed accuracy. So, we proceeded to inspect all single exposure spectra to identify those with a clear emission detection. The seven objects with the lowest S/N were removed from the S2 sample, the remaining 54 objects form the sample S3.

As in our group's previous work, we have selected only those HIIG that have a logarithmic velocity dispersion ($\log \sigma - \epsilon_{\log \sigma} \leq$1.83), which minimises the probability of including rotationally supported systems, leaving 41 objects, sample S4 (see \S \ref{velocity dispersion} for the velocity dispersion measurements). Finally, joining KMOS data with previous ones (obtained with MOSFIRE@Keck and XShooter@VLT, see \S \ref{Repeated observations}), we ended up with 29 new HIIG (S5) that are added to the total sample for the cosmological analysis (see \S \ref{sec:Results} and Table \ref{tab:redshift,sigma and fluxes}). 

%%%%%%%%%%%%%%%%%%% Figure 3 %%%%%%%%%%%%%%%
\begin{figure}
\begin{center}
\includegraphics[width=0.9\columnwidth]{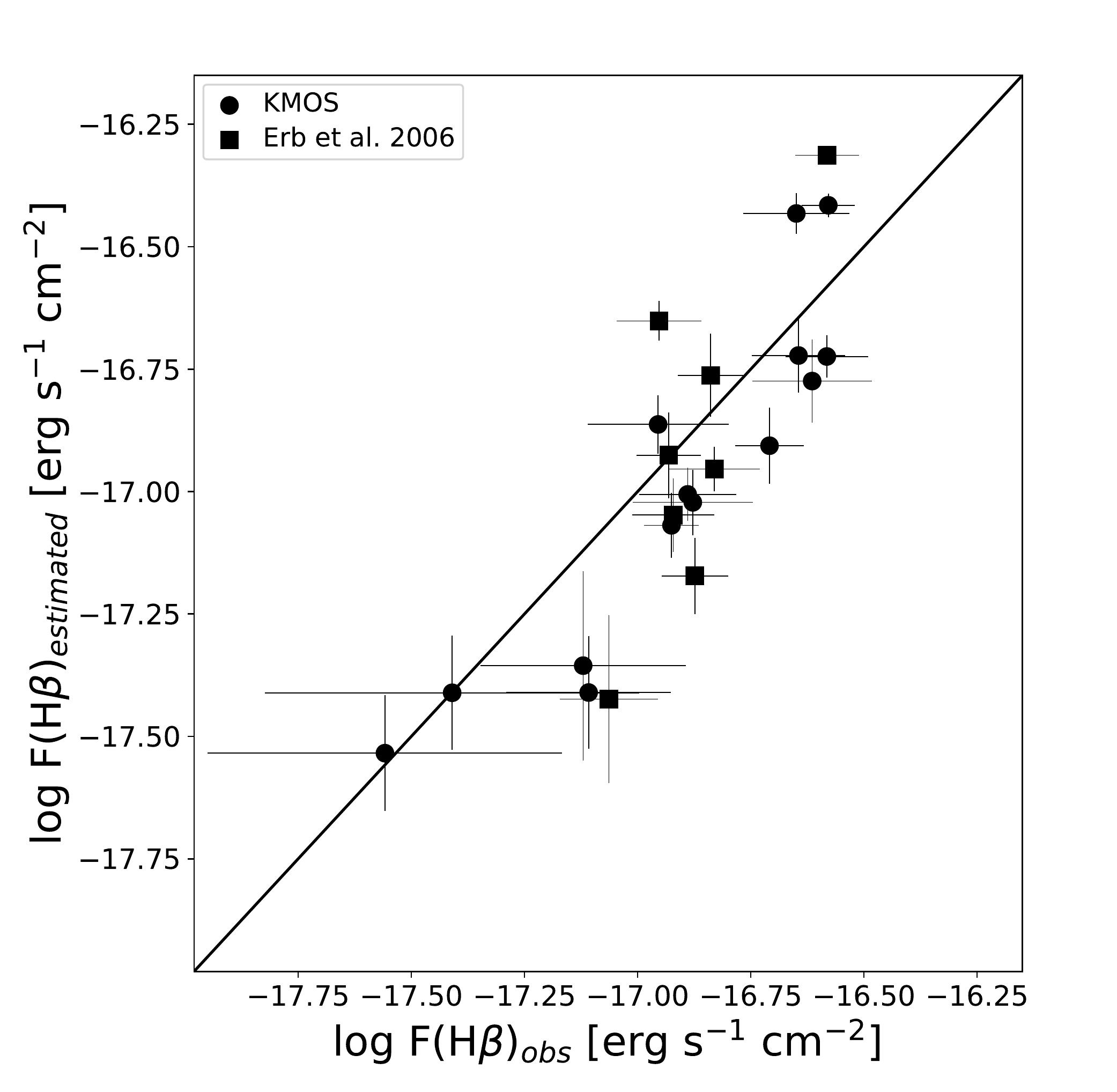}
\end{center}
\caption[FH$\beta$ estimated  vs FH$\beta_{obs}$]{\footnotesize F(H$\beta$) estimated (from F([$\Oiii$]),  see text) versus measured. The  line represents the one-to-one relation.  }
\label{FHb_BPT_vs_FHb_obs}
\end{figure}

\subsection{Emission line widths}
\label{velocity dispersion}

We determine the 1D velocity dispersion ($\sigma$) by fitting a Gaussian profile to each emission line ($\sigma_{obs}$) and subtracting in quadrature the thermal ($\sigma_{th}$), instrumental ($\sigma_i$) and fine structure broadening ($\sigma_{fs}$) components as:

\begin{equation}
 \sigma^2 = \sqrt{\sigma^2_{obs} - \sigma_{th}^2 - \sigma_i^2 - \sigma_{fs}^2}.
\label{eq:sigma}
\end{equation}

The uncertainty in $\sigma$ was estimated using the standard procedure for error propagation with independent errors \citep[see e.g.][]{Wall2012}. The uncertainty in $\sigma_{obs}$ was estimated using a Monte Carlo analysis where a set of random realisations of each spectrum was generated using the r.m.s. intensity of the continuum adjacent to the emission line. The full width at half-maximum of the emission line (FWHM) 1$\sigma$ uncertainty was estimated from the standard deviation of the distribution of FWHM measurements.

Fig. \ref{fig:montecarlo analysis} shows an example of the fit to the observed H$\alpha$ line and the distribution of FWHM obtained from the Monte Carlo simulations (in the inset)  for the target Q2343-BM133. The residuals are shown in the bottom panel.

For the thermal broadening and the fine structure width we adopted the same values as in \cite{GonzalezMoran2019}. The thermal broadening was calculated assuming a Maxwellian velocity distribution of the hydrogen and oxygen ions for which a reasonable value for the electron temperature ($T_e$) is $T_e= 10, 000$ K. The
 instrumental broadening was measured from the width of unsaturated and unblended sky lines.
 
Due to differences of instrumental resolution in each IFU and along the wavelength axis \citep{Davies2013}, there is a variation of the spectral resolution across all 24  IFUs. A unique value for the resolution is derived for each object depending on which IFU and at which wavelength the lines are observed.

We chose to perform the measurements inside the science IFU (instead of the sky one)  because our targets are  point-like  sources (FWHM $<$ 1\arcsec) compared to the IFU FoV (2.8\arcsec $\times$ 2.8\arcsec). Hence the instrumental resolution is derived from the mean FWHM  of OH sky emissions surrounding the line of interest.
The  instrumental resolution uncertainty was calculated using [max(FWHM)$-$min(FWHM)]$/2\sqrt{N}$, where max(FWHM) and min(FWHM) are the  maximum and minimum FWHM of the set of measured sky lines and $N$ is the number of measurements. The  instrumental resolution derived for each object is listed in column 6 of Table \ref{tab:Observational data}.

Line widths are measured from either H$\alpha$ or H$\beta$ for the sample at z $<$ 2, and from [$\Oiii$]$\lambda$5007\AA\ for the objects at z $>$ 2. Several groups \citep[e.g.][]{Hippelein1986, Bordalo_Telles2011, Bresolin2020} have found that the Balmer lines in HIIG and GHIIR are systematically broader than the [$\Oiii$]$\lambda$5007 \AA. In order to bring both measurements into a single system, we corrected the [$\Oiii$]$\lambda$5007\AA\ velocity dispersion measurements using the relation $\sigma$(H$\alpha$) = $\sigma$[$\Oiii$] + (2.91 $\pm$ 0.31) km s$^{-1}$ determined for the local sample of HIIG from \citet{Chavez2014}.

\subsection{Fluxes}
\label{sec:Fluxes}

The emission line fluxes and EW were measured using the IRAF\footnote{IRAF is distributed by the National Optical Astronomy Observatories, which are operated by the Association of Universities for Research in Astronomy, Inc., under cooperative agreement with the National Science Foundation.} task \textit{splot} and their uncertainties were estimated from the  usual expressions \citep[see e.g.][]{Tresse1999} as in  \citet{GonzalezMoran2019}. Due to  sky variations in the H band,  a slightly negative residual background can appear after sky subtraction in the final spectra which needs to be corrected for.  The correction consists of a constant offset  to set to zero the residual background levels in the final 1D spectra. The background levels are estimated from the r.m.s continuum taken from windows that are clean of sky lines. The difference in flux with and without this correction is added as an uncertainty to the estimated flux as:

\begin{equation}
	\epsilon_{Flux} = \sqrt{\epsilon^2_F + \epsilon^2_{back F}},
	\label{eq:sigma_Flux}
\end{equation}
where  $\epsilon_{F}$ was calculated following \citet{Tresse1999} and $\epsilon_{back F}$ is the uncertainty due to the residual background level.

Thirteen objects show only [$\Oiii$]$\lambda$5007\AA\  in the observed spectrum. For 9 of these  \citet{Erb2006a} published the value of  
F(H$\alpha$) and we calculated F(H$\beta$) from the theoretical F(H$\alpha$)/F(H$\beta$) ratio (2.86) expected for Case B recombination \citep{Osterbrock1989} with $T_e=10,000$ K and low densities (N$_e<100$ cm$^{-3}$) and assuming the mean extinction that will be discussed in \S \ref{sec:Extinction correction}. For the remaining 4 objects, the F([$\Oiii$]) was transformed to F(H$\beta$) using the mean ratio F([$\Oiii$])/F(H$\beta$) obtained for our local sample of HIIG \citep{Chavez2014}.

In order to check the reliability of this method, we compared the measured F(H$\beta$) with F([$\Oiii$]) for those objects with both measurements, and the comparison is shown in Fig. \ref{FHb_BPT_vs_FHb_obs}. Based on this comparison, we kept the four objects for which we have only F([$\Oiii$]) in the analysis.

%%%%%%%%%%%%%%%%% Figure 4 %%%%%%%%%%%%%
\begin{figure*}
%\vspace*{-0.3cm}
%\begin{center}
\includegraphics[width=1.0\textwidth]{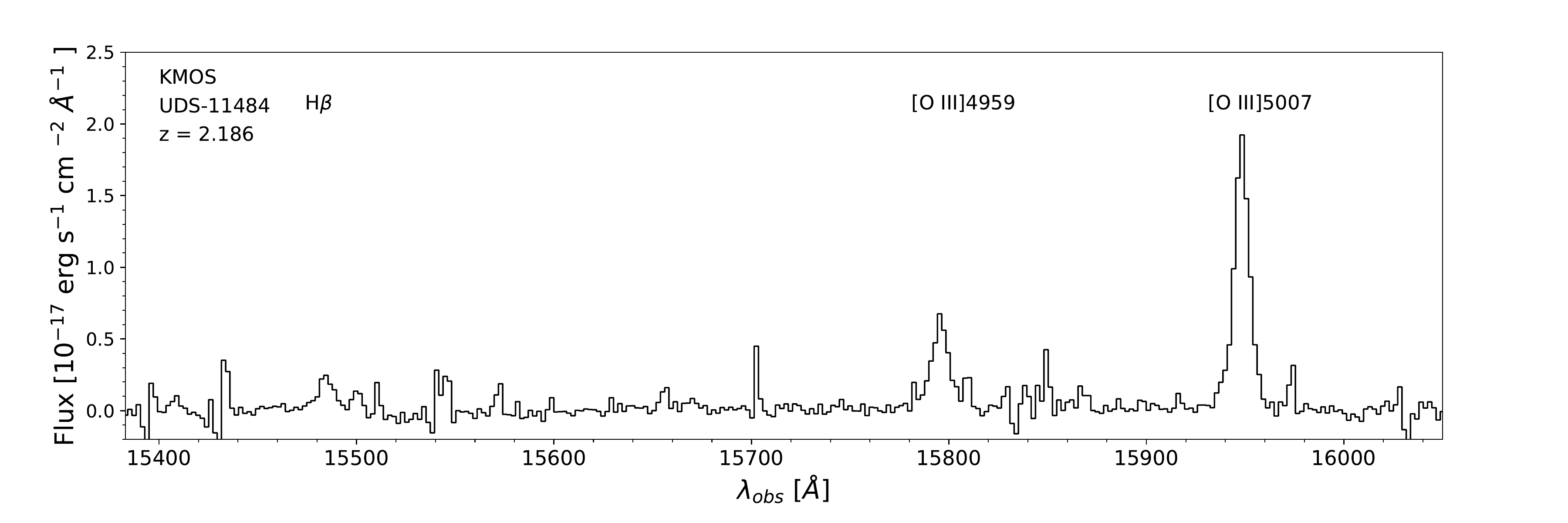}
%\end{center}

%\vspace*{-0.7cm}
%\begin{center}
\includegraphics[width=1.0\textwidth]{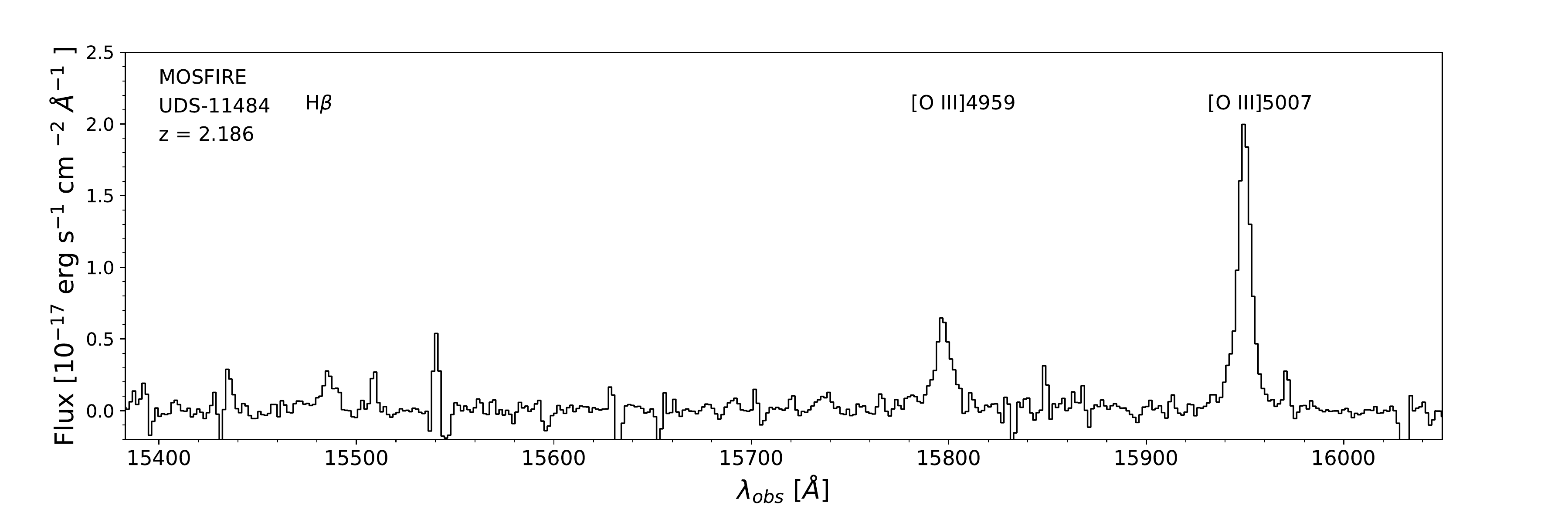}
%\end{center}
\caption{UDS-11484 1D spectra. Top panel: KMOS observations with a total exposure time of 9,400 seconds. Bottom panel: MOSFIRE observations with a total exposure time of 7,200 seconds. Both spectra have similar S/N per unit time and unit wavelength.}
\label{fig:comparison for UDS-11484}
\end{figure*}

\subsection{Repeated observations: KMOS, MOSFIRE, XShooter }
\label{Repeated observations}

With the intention of linking data obtained at different sites, with different instruments and telescopes, we observed with KMOS some objects from  our previous samples.
Twelve  objects had previously been observed with MOSFIRE@Keck  and XShooter@VLT with higher spectral resolution in the H band [R(MOSFIRE) $\sim$ 5400 and R(XShooter) $\sim$ 8000 against  R(KMOS) $\sim$ 4000]. We chose to use the velocity dispersion for these objects from the higher dispersion data published in \citet[][for XShooter]{Terlevich2015} and \citet[][for MOSFIRE]{GonzalezMoran2019}.

The nine  objects in common with MOSFIRE are UDS23, UDS25, UDS40, UDS-11484, UDS-14655, UDS-4501, COSMOS-16566, COSMOS-18358 and  zCOSMOS-411737. In Fig. \ref{fig:comparison for UDS-11484} we show an example of the spectra obtained for the same object (UDS-11484 at z=2.186) with KMOS and MOSFIRE in the region covering [$\Oiii$]$\lambda\lambda$4959, 5007\AA\ and H$\beta$. Both spectra have similar S/N per unit time and unit wavelength.
The three targets previously observed with XShooter are Q2343-BM133, Q2343-BX418 and Q2343-BX660. 

\begin{figure}
\begin{center}
\includegraphics[width=0.9\columnwidth]{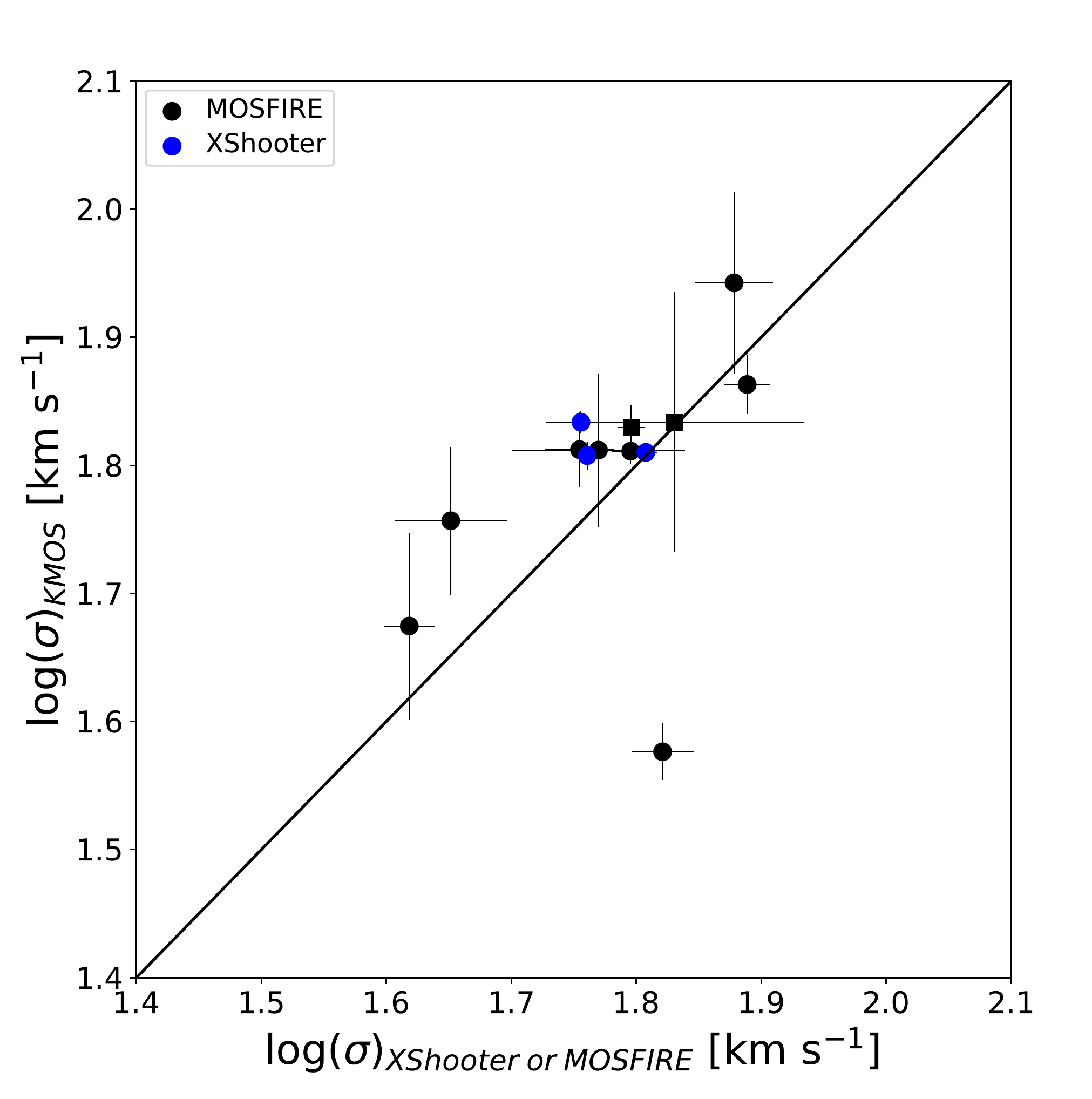}
\end{center}
\caption{Comparison between the velocity dispersion from MOSFIRE or XShooter and KMOS data. The solid line represents the one to one relation. The two squares correspond to UDS-11484 for the velocity dispersion determined from either H$\beta$ or [O III]$\lambda$5007\AA. The largest error bars correspond to the H$\beta$ value. The outlier is COSMOS-18358.}
\label{fig:sigmas_comparison}
\end{figure}

Fig. \ref{fig:sigmas_comparison} shows the comparison between the velocity dispersion determined from either H$\alpha$, H$\beta$ or [O III]$\lambda$5007\AA\ for the targets with repeated observations in KMOS and XShooter (blue circles) or in KMOS and MOSFIRE (black circles). The target UDS-11484 at z=2.2 appears twice (black squares) as its $\sigma$ was measured from both H$\beta$ and [$\Oiii$].

%%%%%%%%%%%%%%%%%% Figure  5 %%%%%%%%%%%%%%%%%%

\begin{figure*}
\begin{center}$
\begin{array}{cc}
\vspace{-1cm}
  \hspace{-1cm} {\includegraphics[width=.6\textwidth]{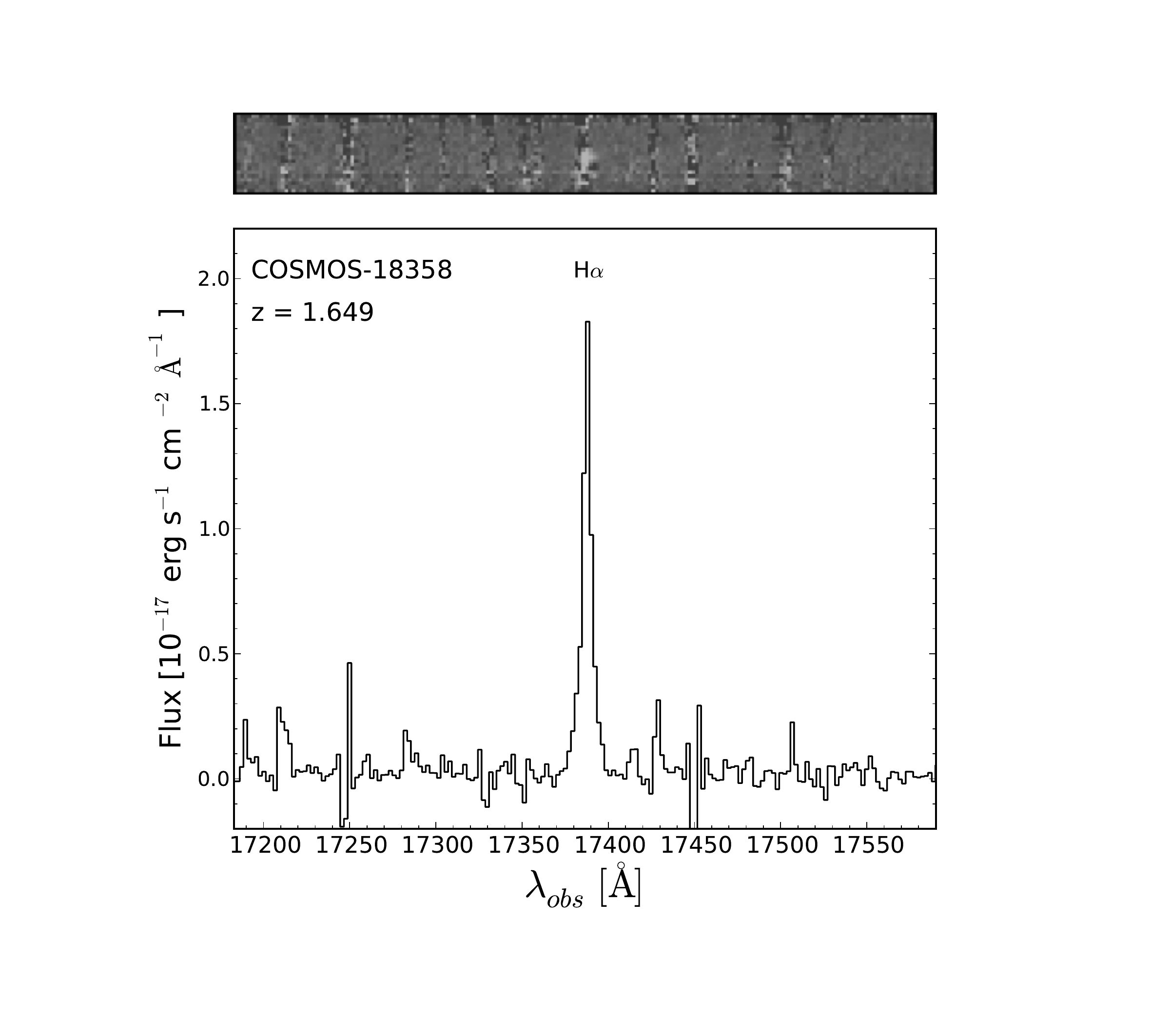}} &
  \hspace{-2cm}{\includegraphics[width=.603\textwidth]{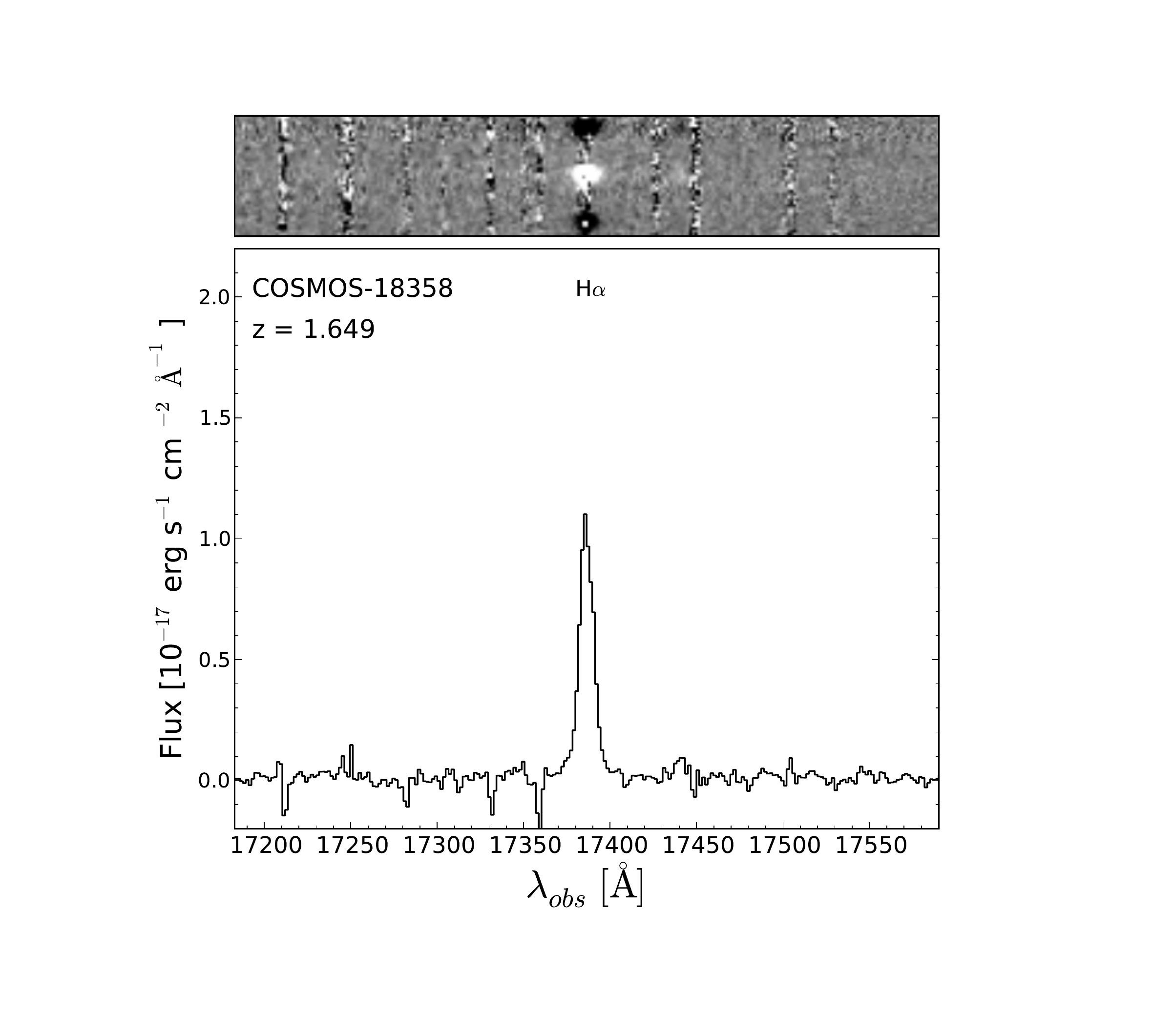}}  
\end{array}$
\end{center}
\caption{COSMOS-18358 2D (top) and 1D (bottom) spectra. Left panel: KMOS data with a total exposure time of 9 800 seconds. Right panel: MOSFIRE data with a total exposure time of 10 080 seconds.}
\label{fig:comparison for COSMOS18358}
\end{figure*}

We can see from Fig. \ref{fig:sigmas_comparison} that the measurements are in agreement within 1$\sigma$ except for the outlier source COSMOS-18358. In order to understand this discrepancy, we compare its  KMOS and MOSFIRE 1D and 2D spectra  (see Fig. \ref{fig:comparison for COSMOS18358}). To extract the 2D KMOS spectrum, a pseudo-longslit from the data cube was created using the function \textit{longslit} in QFitsView. It  is clear from the figure that for this object, the sky subtraction for the MOSFIRE spectrum is better than for the KMOS one for which  a sky emission line  coincides with H$\alpha$. From the redshifts measured in both spectra, we find that z$_{MOSFIRE}$ is slightly lower than z$_{KMOS}$ (1.6492 vs 1.6494) which confirms that the blue side of  H$\alpha$  in the KMOS spectrum is affected by a bad subtraction of a sky emission line. In consequence, considering also the better spectral resolution of the MOSFIRE data, we use in what follows the velocity dispersion calculated from the MOSFIRE spectra whenever possible.

\begin{figure}
\begin{center}
\hspace{-0.3cm}\includegraphics[width=0.92\columnwidth]{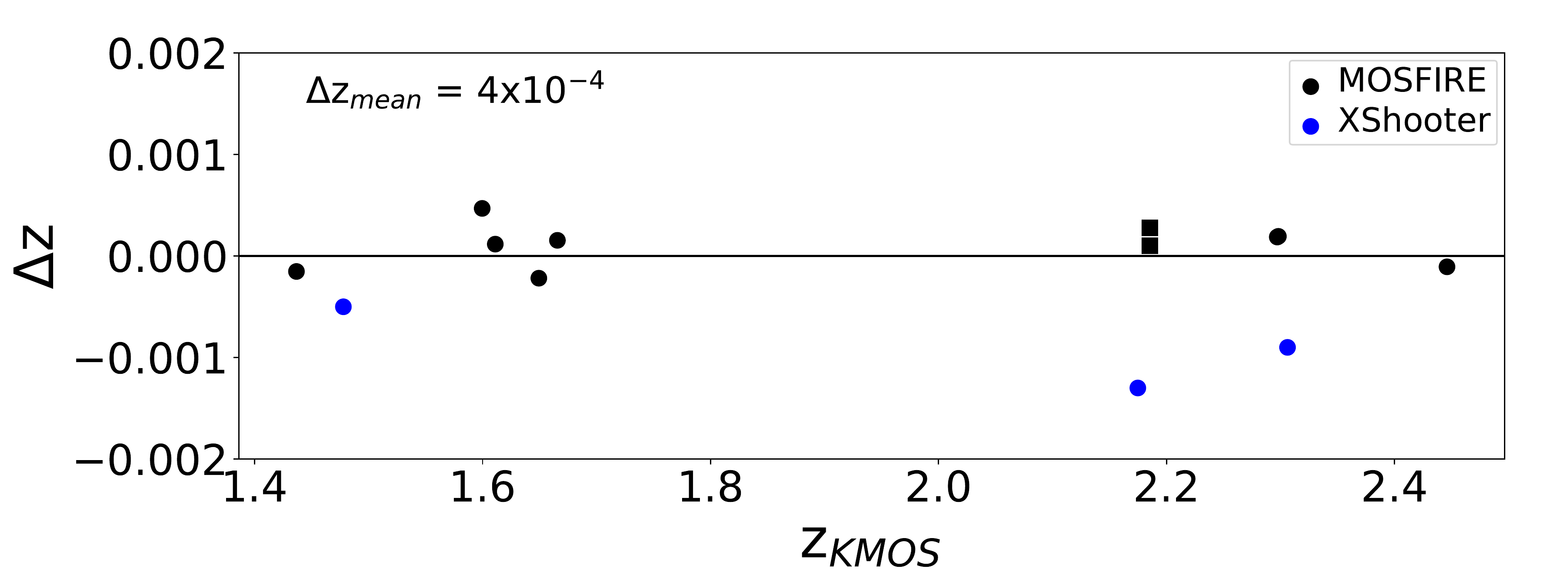}
\end{center}
\caption[Redshift difference for the targets observed with MOSFIRE or XShooter and with KMOS]{\footnotesize $\Delta$z: redshift difference for the targets observed with MOSFIRE or XShooter and with KMOS compared to the value obtained from KMOS data.}
\label{fig:redshift_comparison}
\end{figure}

Fig. \ref{fig:redshift_comparison} shows the redshift difference  $\Delta$z for the  targets with either MOSFIRE or XShooter previous observations.
The mean of $\Delta$z  is 4$\times$10$^{-4}$ and as individual redshift uncertainties given by the Gaussian fit to the emission lines are of the order of 10$^{-5}$, we set the redshift uncertainty for the KMOS and MOSFIRE data to 4$\times$10$^{-4}$  for a more realistic value.

%%%%%%%%%%%%%%%%%% Figure  6 %%%%%%%%%%%%%

%%%%%%%%%%%%%%%%%% Figure  7 %%%%%%%%%%%%%%%%%%

\begin{figure}
\begin{center}
\includegraphics[width=0.9\columnwidth]{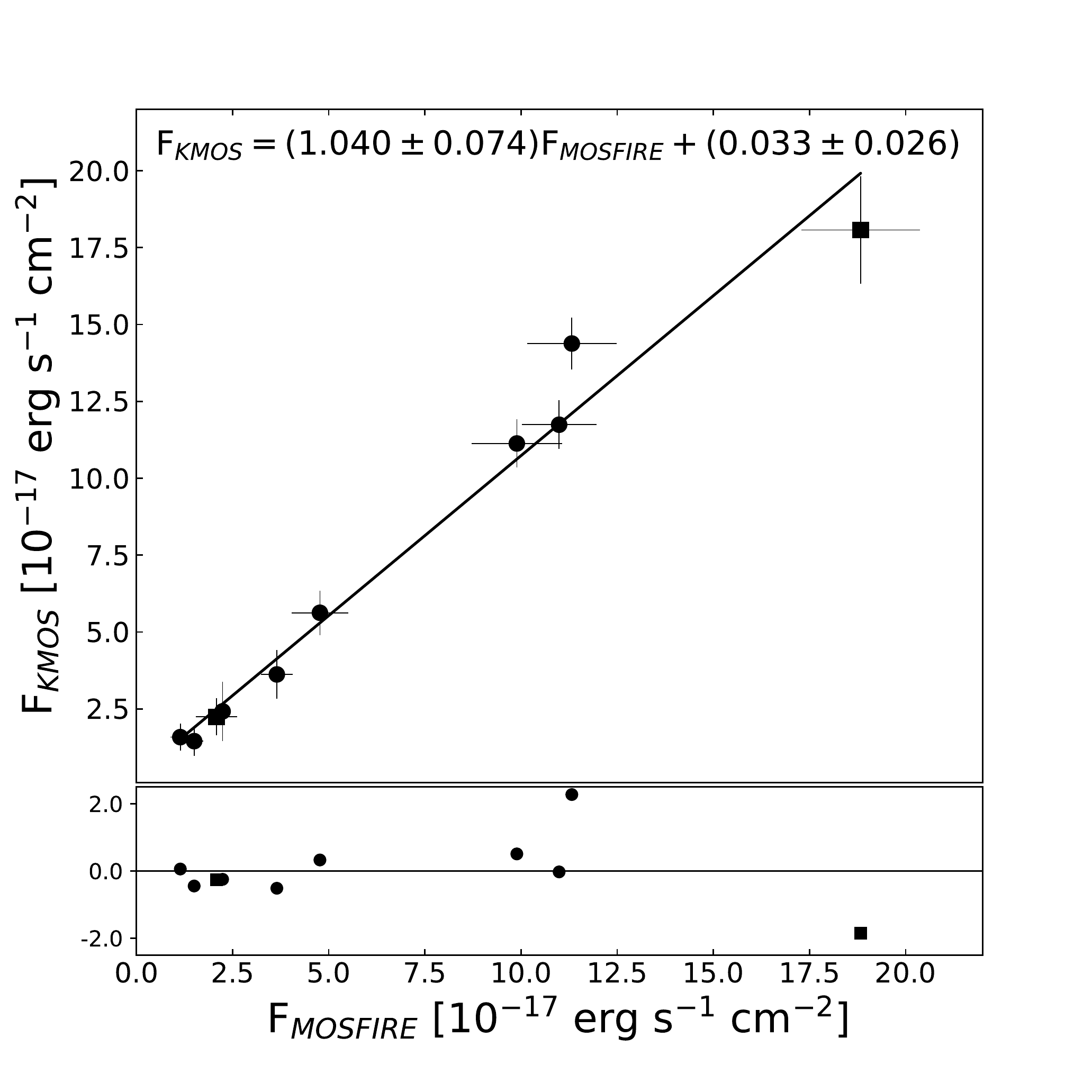}
\end{center}
\caption[MOSFIRE and KMOS fluxes comparison]{\footnotesize Fluxes measured in repeated observations with MOSFIRE vs. KMOS. The black line is the fit and the box underneath shows the residuals.}
\label{MOSFIRE-KMOS fluxes}
\end{figure}

The KMOS IFU FoV (2.8\arcsec $\times$ 2.8\arcsec) ensures that there is virtually no flux loss for the high redshift HIIG (which are smaller than 1\arcsec, see Fig. \ref{KMOS sample images}).
 This was not the case for the MOSFIRE observations \citep{GonzalezMoran2019} where a slit width of 0.48\arcsec was used. A comparison of the flux for the 9 targets observed with both instruments provides an estimate of the slit loss which is henceforth used to correct the fluxes measured from the MOSFIRE data. This is shown in Fig. \ref{MOSFIRE-KMOS fluxes} where the continuous line is the linear fit obtained using the mpfit routine. Note that, as in Fig. \ref{fig:sigmas_comparison}, there are 10 points instead of 9 in the figure for the two velocity dispersion values obtained using H$\beta$ and [$\Oiii$]$\lambda$5007\AA\ in UDS-11484. Even though there is not much difference between MOSFIRE and KMOS fluxes, we corrected the MOSFIRE sample fluxes for slit loss. The results are shown in Table \ref{tab:redshift,sigma and fluxes}.

%%%%%%%%%%%%%%%%%% Figure  8 %%%%%%%%%%%%%%%%%%

\subsection{Extinction correction}
\label{sec:Extinction correction}
Extinction correction was determined using the \citet[][]{Gordon2003} extinction law chosen because the dust attenuation curves derived from analogs of high redshift star forming galaxies by \citet{Salim2018} and from star forming galaxies at z $\sim$ 2 by \citet{Reddy2015} are in good agreement with the LMC and SMC curves given by \citet{Gordon2003} \citep[see Fig.  3 from][]{GonzalezMoran2019}.

The extinction corrected fluxes were determined as  usual  from the expression:
\begin{equation}
F(\lambda) = F_{obs}(\lambda)10^{0.4\textsc{Av}k(\lambda)/\textsc{Rv}},
\label{eq:flux correction}
\end{equation}
where $k(\lambda) = A(\lambda)/E(B - V )$ is given by the extinction law used. We adopt $k(H\beta) = 3.33$ and $k(H\alpha) = 2.22$ and Rv = 2.77  \citep{Gordon2003}.

Given the objects redshift, we cannot measure the Balmer decrement  from the H band data. We adopted instead the mean extinction (Av=0.71$\pm$0.13) derived for our  local sample \citep{Chavez2014}.

%%%%%%%%%%%%%%%%%%%% Figure  9 %%%%%%%%%%%%%%%
\begin{figure}
\begin{center}
\includegraphics[width=1.0\columnwidth]{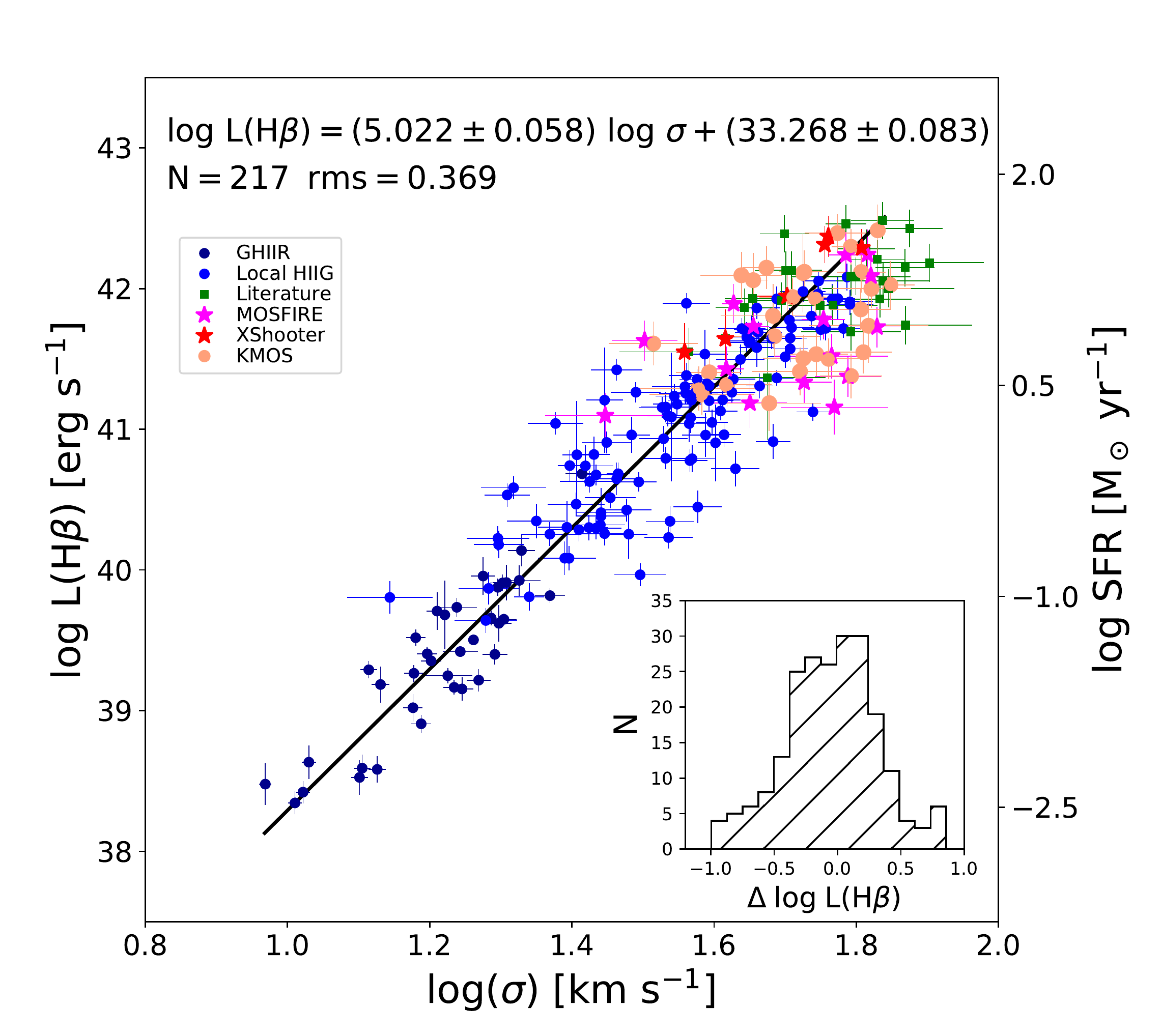}
\end{center}
\caption[L-sigma relation.]{\footnotesize \lsigG\ relation for the Global sample (see Table \ref{tab:Samples used in the HIIG cosmological analysis}) using the \citet{Gordon2003} extinction curve. The right axis shows the SFR estimated from the L(H$\beta$) using SFR$=1.54\times10^{41}$L(H$\beta$). The black line is the fit, obtained only from the local samples, shown at the top and the lower right corner box  shows the residuals.}
\label{fig:KMOS-MOSFIRE L-sigma}
\end{figure}

\section{Results}
\label{sec:Results}

For  cosmological parameter analysis we include both our new KMOS data and previously published samples of low- and high-z targets. This is summarised in Table \ref{tab:Samples used in the HIIG cosmological analysis}, where the first column gives the reference name of the sample, the second lists its description and the
third gives the number of objects in each subsample.

\begin{figure*}
\hspace*{-0.7cm}
%\begin{center}
\includegraphics[width=2.2\columnwidth]{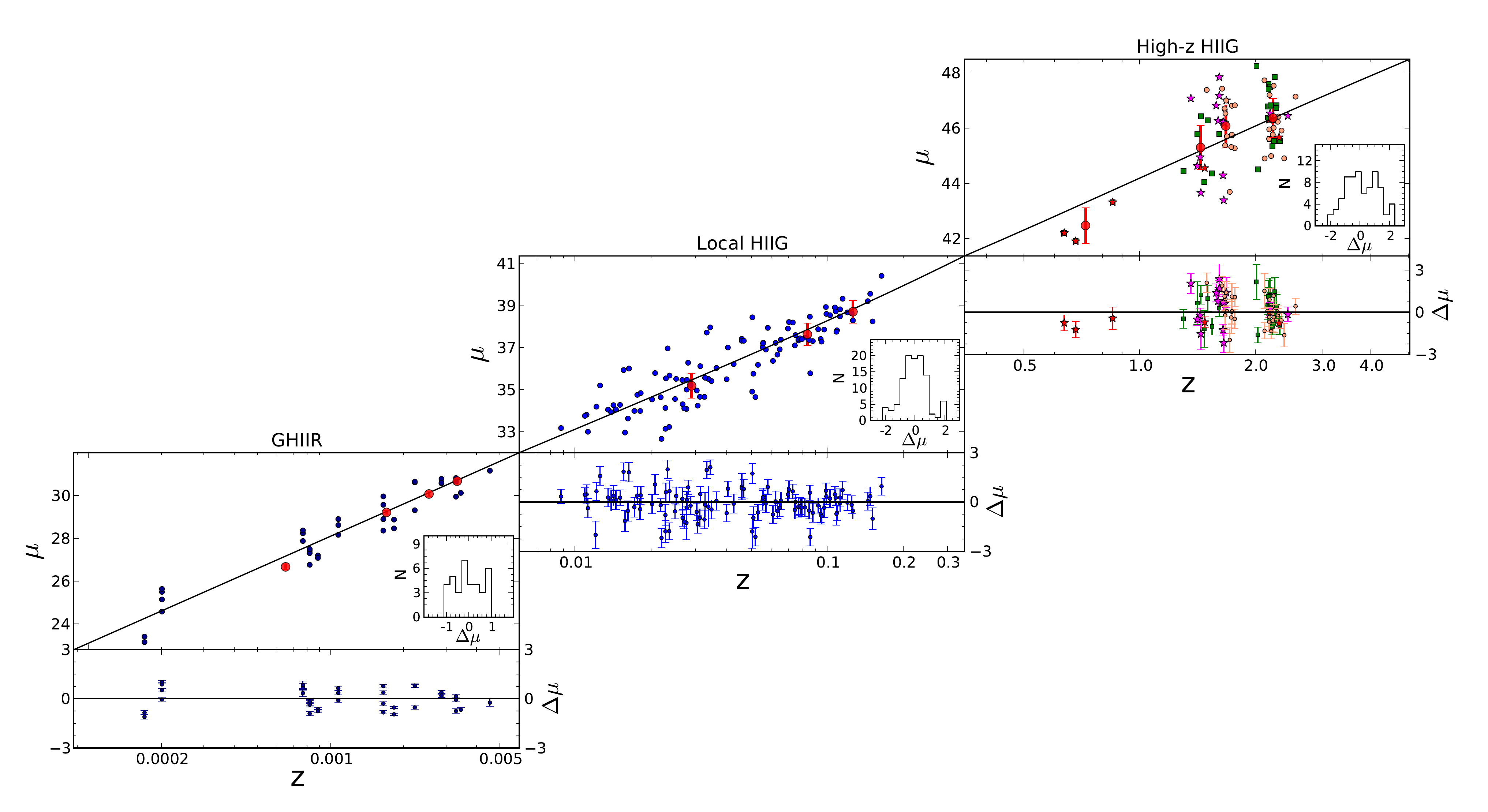}
%\end{center}
\caption[KMOS-MOSFIRE Hubble Diagram.]{\footnotesize Hubble diagram connecting our local and high redshift samples up to z $\sim$ 2.6. Red circles represent averages of the distance moduli in redshift bins; the rest of the symbols  are as in Fig. \ref{fig:KMOS-MOSFIRE L-sigma}. The continuous line corresponds to $\Omega_m=0.249$ and $w_0=-1.18$ (our best cosmological model using only HIIG, see \S \ref{sec:ConstrainingDEEoS}). The insets show the distribution of the residuals of the fit that are plotted in the bottom panels.}
\label{fig:KMOS-MOSFIRE Hubble Diagram}
\end{figure*}

The \lsigG\ relation shown in Fig. \ref{fig:KMOS-MOSFIRE L-sigma} includes the new data for high-z objects presented in \S \ref{sec:KMOS Observations}, in addition to the sample in \citet{GonzalezMoran2019}, giving a total of 74 high-z HIIG.
The intercept ($\alpha$) and slope ($\beta$) of the relation, shown in the figure inset, are estimated for the sample of 107 local  ($0.01 \leq z \leq 0.15$) HIIG published in \citet{Chavez2014} and 36 GHIIR at z $\leq 0.01$ described in \cite{Fernandez2018} using the \citet{Gordon2003} extinction curve. We use the values of $\alpha$ and $\beta$ so obtained as the nuisance parameters of the \lsigG\ relation unless stated otherwise. Fig. \ref{fig:KMOS-MOSFIRE Hubble Diagram} shows the Hubble diagram for the Global sample of 217 objects at $0.01 \leq  z \leq 2.6$ (see Table \ref{tab:Samples used in the HIIG cosmological analysis}) where  the redshift range covered by our data can be clearly appreciated. 

%%%%%%%%%%%%%%%%%%% Table %%%%%%%%%%%%%%
\begin{table}
\caption{Samples used in the  cosmological analysis.}
\label{tab:Samples used in the HIIG cosmological analysis}
%\centering
\resizebox{\columnwidth}{!}{
\begin{tabular}{ l l l }
\hline \hline
Sample & Description & N \\
\hline
KMOS   &  S5 sample   & 29 \\
MOSFIRE  &  MOSFIRE sample corrected by slit loss flux &  15 \\
XShooter   &   XShooter sample corrected by slit loss flux   &  6  \\
Literature   &  Literature sample$^a$      & 24 \\
High-z  & KMOS + MOSFIRE + XShooter + Literature & 74 \\
Local & Local HIIG sample & 107  \\
Full & High-z + Local  & 181 \\
Our data  & Full excluding Literature & 157 \\
GHIIR & GHIIR sample & 36 \\
Global & Full + GHIIR & 217 \\
\hline
\hline
\multicolumn{3}{l}{}\\
\multicolumn{3}{l}{$^a$ \citet{Erb2006a, Masters2014} and \citet{Maseda2014}.}
\end{tabular}}
\end{table}

\subsection{Cosmological parameters constraints}

To constrain cosmological parameters in a way that is independent of  $h$, we follow the methodology from our previous work \citep{Chavez2016, GonzalezMoran2019} which we summarise in what follows. 

The likelihood function for HIIG and GHIIR  is given as:
\begin{equation}
        \mathcal{L}_{HII} \propto \exp{(- \frac{1}{2} \chi^2_{HII})},
\label{eq:lkh}
\end{equation}
where:
\begin{equation}
        \chi^2_{HII} =  \sum_{n}\frac{(\mu_o(\log f, \log \sigma | \alpha, \beta) - \mu_{\theta}(z | \theta))^2}{\epsilon^2},
\label{eq:chi}
\end{equation}
and $\mu_o$ is the distance modulus calculated from the observables, $\mu_{\theta}$  is the theoretical distance modulus, $\alpha$ and $\beta$ are the \lsigG\ relation's intercept and slope respectively; $ \sigma$ is the broadening corrected velocity dispersion and $f$ is the extinction corrected  flux.

%%%%%%%%%%%%%%%%%%% Figure 10 HubbleDiagram %%%%%%%%%%%%%%%%

The theoretical distance modulus $\mu_{\theta}$ in Eq. \ref{eq:chi}  depends on a set of cosmological parameters, in the most general case considered here given as  $\theta = \{h, \Omega_m, w_0, w_a \}$,  and the redshift ($z$). The parameters $w_0$ and $w_a$ refer to the DE EoS, the general form of which is:
\begin{equation}
p_{w} =  w(z) \rho_{w} c^2 \;,
\end{equation}
with $p_{w}$ the pressure and $\rho_{w}$ the density of DE, while $w(z)$ is an evolving DE EoS parameter. There are different DE models, many are parametrized using a Taylor expansion around the present epoch like the CPL model \citep{Chevallier2001, Linder2003, Peebles2003, Dicus2004, Wang2006} in which:
\begin{equation}
        w(z)=w_0+w_a \frac{z}{1 + z}\;,
\end{equation}

The cosmological constant is just a special case of DE, given for $(w_0,w_a)=(-1,0)$, while the so called wCDM models are such that $w_a=0$ but $w_0$ can take values $\neq -1$.

Finally $\epsilon^2$, the weights in the likelihood function,  can be given as:
  \begin{equation}
	  \epsilon^2 = \epsilon^2_{\mu_{o}, stat} + \epsilon^2_{\mu_{\theta}, stat} + \epsilon^2_{sys},
\label{eq:epsilon}         
 \end{equation}
 where $\epsilon_{\mu_{o}, stat}$ are the statistical uncertainties  given as:
 \begin{equation}
	 \epsilon^2_{\mu_{0}, stat} = 6.25(\epsilon_{\log f}^2 + \beta^2\epsilon_{\log \sigma}^2 + \epsilon_{\beta}^2\log \sigma^2 + \epsilon_{\alpha}^2).
  \end{equation}
$\epsilon_{\log f}$,  $\epsilon_{\log \sigma}$, $\epsilon_{\alpha}$  and $\epsilon_\beta$ are the uncertainties associated with the logarithm of the flux, the logarithm of the velocity dispersion and the intercept and slope of the \lsigG\ relation respectively, while $\epsilon_{\mu_{\theta}, stat}$  in Eq. \ref{eq:epsilon} is the uncertainty associated with the distance modulus as propagated from the redshift uncertainty in the case of HIIG and  as given by the primary distance indicator measurement uncertainty for the case of GHIIR. $\epsilon_{sys}$ are the systematic uncertainties that will be briefly discussed in \S \ref{sec:Discussion} and in more detail in a forthcoming paper (Ch\'avez et al., in prep).

It is convenient to define also an $h$-free likelihood function \citep[cf.][]{Nesseris2005} through a rescaling of the luminosity distance ($d_L$) given by:
\begin{equation}
D_L(z,\theta)= (1+z) \int_{0}^{z}{\frac{dz^{'}}{E(z^{'}, \theta)}}\;,
\end{equation}
i.e., $d_L=c D_L/H_0$. We use this rescaling to constrain cosmological parameters in an $h$ independent way as fully described in \citet{GonzalezMoran2019}.

The \lsigG\ relation has been used in the local Universe to constrain the value of $h$ \citep{Chavez2012,Fernandez2018}. The main objective of this work is to constrain the parameters $\theta = \{\Omega_m, w_0, w_a \}$ in a way that is independent of $h$. 
However, we will also constrain $h$ in some cases, using the full likelihood function as given in Eq. \ref{eq:lkh} in order to  compare our results with the literature. 
For each result we will specify whether or not it is independent of $h$ and which parameters have been left fixed.

Unless otherwise stated, we use the MultiNest Bayesian inference algorithm \citep[cf.][]{Feroz2008, Feroz2009, Feroz2013}, to maximise the likelihood function and get constraints to the different combinations of nuisance and cosmological parameters. In all the cases we use the priors given in Table \ref{tab:priors} \citep{Chavez2016}.

%%%%%%%%%%%%%%%%%%% Table 3 %%%%%%%%%%%%%%
\begin{table}
\caption{Priors for Constrained Parameters.}
\label{tab:priors}
\centering
\begin{tabular}{ l l }
\hline \hline
Parameter & Prior \\
\hline
\multicolumn{2}{c}{Cosmological Parameters}\\
\hline
$h$ & Uniform [0.5, 1.0] \\
$\Omega_m$ & Uniform [0.0, 1.0] \\
$w_0$ & Uniform [$-$2.0, 0.0] \\
$w_a$ & Uniform [$-$4.0, 2.0] \\
$w_b$ & Uniform [0.0, 0.05] \\
\hline
\multicolumn{2}{c}{HIIG Nuisance Parameters}\\
\hline
$\alpha$ & Uniform [32.5, 34.5] \\
$\beta$ & Uniform [4.5, 5.5] \\
\hline
\hline
\end{tabular}
\end{table}

\subsubsection{Constraining $\Omega_m$}
\label{sec:ConstrainingOm}
Applying the $h$ independent method described above to the joint local and high-$z$ sample of 181 HIIG (dubbed the Full sample in Table \ref{tab:Samples used in the HIIG cosmological analysis}), and assuming the standard $\Lambda$CDM model with $w_0 = -1$, we find  $\Omega_m = 0.236^{+0.047}_{-0.041}$ (stat) and $\Omega_m = 0.244^{+0.040}_{-0.049}$ (stat) for the $\chi^2$-minimisation procedure and the MultiNest MCMC, respectively. The last result is also shown in Table \ref{tab:parLit}. If we restrict the sample to the 157 HIIG observed by our group (i.e excluding the data taken from the literature, Our data in Table  \ref{tab:Samples used in the HIIG cosmological analysis}), we obtain $\Omega_m = 0.243^{+0.047}_{-0.057}$ (stat); the posterior for this case is shown in Table \ref{tab:par}.

%%%%%%%%%%%%%%%%%%% Figure  11 JOm %%%%%%%%%%%%%%%%%%%%%%%
\begin{figure}
\begin{center}
\includegraphics[width=1.0\columnwidth]{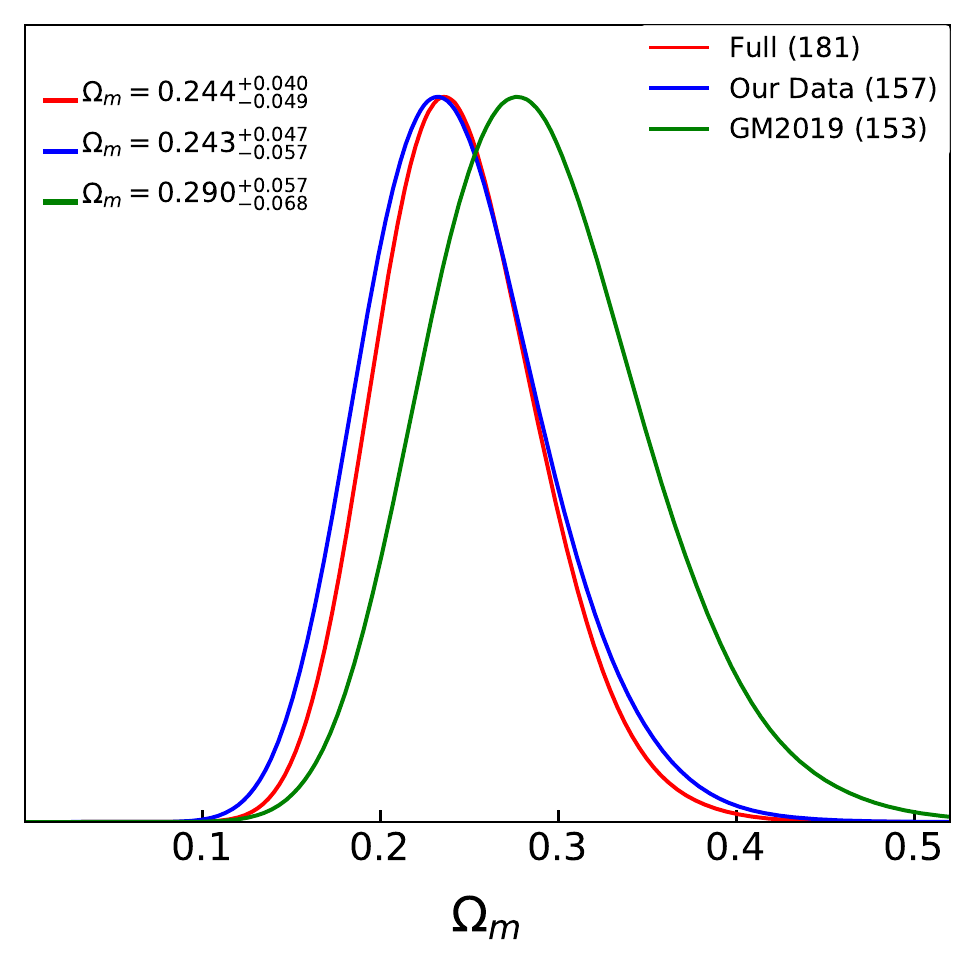}
\end{center}
	\caption{Comparison between the $\Omega_m$ posterior for our current HIIG sample with and without data from the literature (Full and Our Data, respectively) and that of \citet[][GM2019]{GonzalezMoran2019}. The number of objects in each sample is shown between parentheses.}
\label{fig:JOm}
\end{figure}

%When we exclude the 24 high-z objects  from the literature and use only our own sample of 159 HIIG  we find \ojo\ ???

For the purpose of comparing our current constraints with other determinations, we adopt the definition of figure of merit ($FoM$) given by \citet{Wang2008}:
\begin{equation}
	FoM = \frac{1}{\sqrt{det\ Cov(\theta_0, \theta_1, \theta_2, ...)}} \,
\end{equation}
where $Cov(\theta_0, \theta_1, \theta_2, ...)$ is the covariance matrix of a set of parameters $\{\theta_i\}$. 

Our current constraints can be compared with the value of $\Omega_m=0.290^{+0.056}_{-0.069}$ (stat) ($FoM = 16.01$) for the sample of 153 objects  \citep{GonzalezMoran2019}, which includes the Literature sample. We find an improvement of the cosmological parameter constraints by 21\% using Our data sample ($FoM = 19.38$) or by 37\% when considering the Full sample ($FoM = 22.0$). We also compare the posteriors for these three different cases in Fig. \ref{fig:JOm}.

\begin{figure}
\begin{center}
\includegraphics[width=1.0\columnwidth]{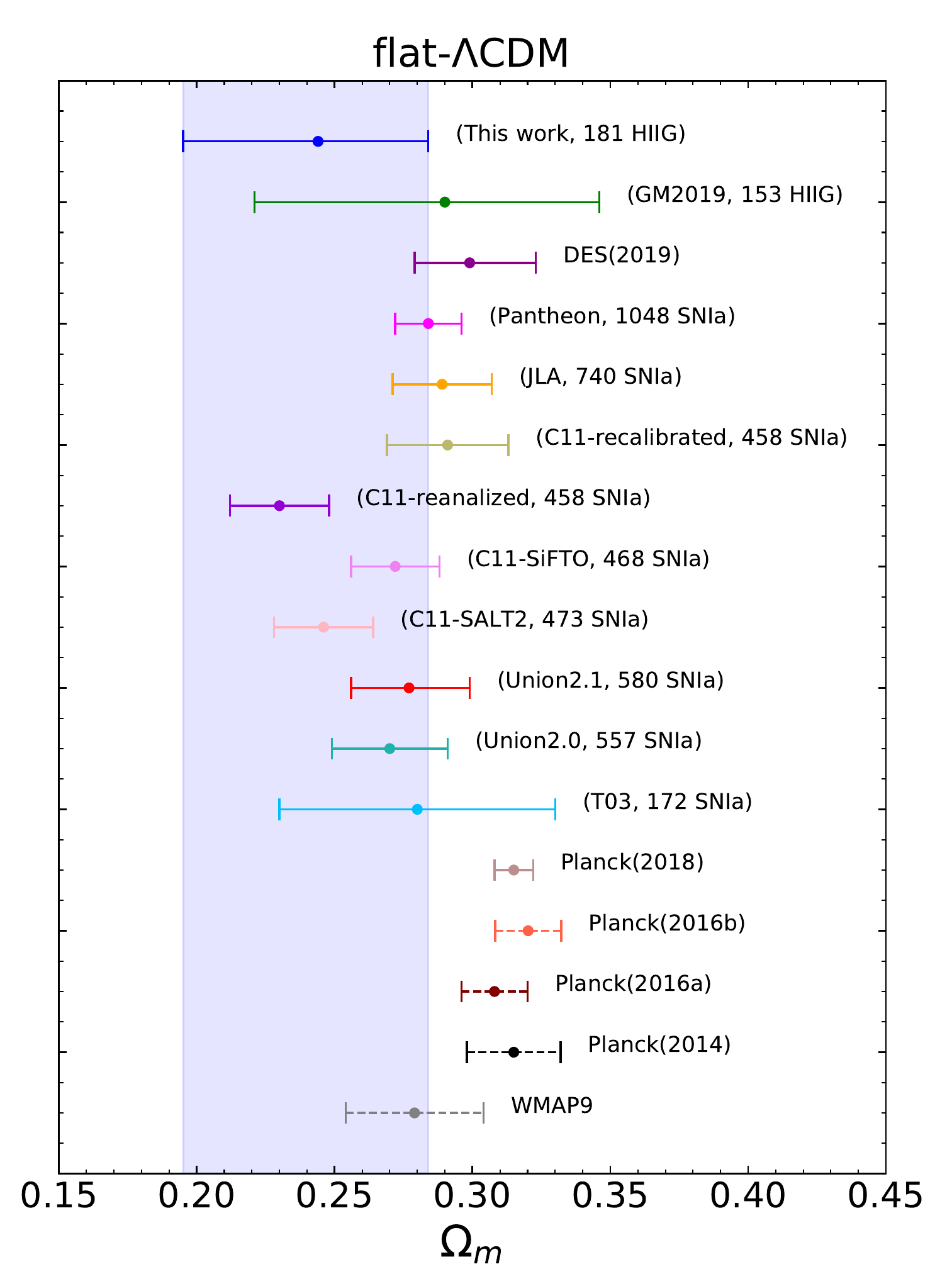}
\end{center}
        \caption{Comparison of our $\Omega_m$ value for the Full sample with literature results. The number of objects and the distance estimator in each sample are shown in parentheses. The continuous error bars are the statistical uncertainties; dashed ones represent both statistical plus systematic uncertainties. The shaded region represents the uncertainties of our current $\Omega_m$ value.}
\label{fig:valoresOM}
\end{figure}

The comparison of our value for  $\Omega_m$ with other  results from the literature is summarised in Fig. \ref{fig:valoresOM} where  dashed error bars denote statistical and systematic uncertainties and  continuous error bars, only statistical. The shaded region (given as a visual aid) represents the uncertainties of our current $\Omega_m$ value.

%%%%%%%%%%%%%%%%%%% Figure 12  Om comparison %%%%%%%%%%%%%%%%%%%%%%

Our result is in agreement with the CMB determination from WMAP9 \citep{Bennett2013} of $\Omega_m=0.279 \pm 0.025$ at a $<1\sigma$ level. It is in disagreement with those found by \citet{PlanckCollaboration2016a, PlanckCollaboration2016b,PlanckCollaboration2018} of $\Omega_m=0.308 \pm 0.012$, $\Omega_m= 0.320 \pm 0.009$ (stat) and $\Omega_m= 0.315 \pm 0.007$ (stat + sys) by 1.4, 1.7 and 1.6$\sigma$, respectively, although we are only considering random errors and the Planck collaboration included systematics.

Our determination of $\Omega_m$ is also in agreement with the result from the Dark Energy Survey (DES) collaboration of $\Omega_m=0.299^{+0.024}_{-0.020}$ \citep{Abbott2019} which results from the joint analysis of BAO, SNIa, weak lensing and galaxy clustering.

As systematic uncertainties for the HIIG sample are not included in the present analysis, it is interesting to compare with the results presented by \citet{Betoule2014}, who analyse the drift on the value of $\Omega_m$ (and SNIa analysis nuisance parameters) with respect to the results presented by \citet[hereafter C11]{Conley2011},  considering statistical uncertainties only. They found $\Omega_m=0.246 \pm 0.018$ (stat) and $\Omega_m=0.272 \pm 0.016$ (stat), for the C11 sample analysis using the SALT2 \citep{Guy2007} and SiFTO \citep{Conley2008} light-curve models respectively. \citet{Betoule2014} recalibrated the C11 sample and found $\Omega_m=0.291 \pm 0.022$ (stat) for 458 SNIa which is in agreement with their value of  $\Omega_m=0.289 \pm 0.018$ (stat) for the JLA sample of 740 SNIa. Our determination is  compatible with both results.
%A comprehensive discussion  of the systematic errors for our HIIG work is the subject of a forthcoming paper.

As can be seen in Fig.  \ref{fig:valoresOM}, the size of the error bars for the SNIa based $\Omega_m$ largely depends on the number of targets used. For example, the High-z Supernova Search Team with 172 SNIa \citep{Tonry2003} found $\Omega_m=0.28 \pm 0.05$ (stat). From this result, we can conclude that for a comparable size of the sample used,  our uncertainties are similar to those from SNIa. The two samples however span a different redshift range: up to 1.2 for \citet{Tonry2003} and up to 2.6 for ours. We aim at filling the gap seen between 0.2 $<$ z $<$ 1.2 in our Hubble diagram (Fig.  \ref{fig:KMOS-MOSFIRE Hubble Diagram}) with forthcoming ground based observations.

%%%%%%%%%%%%%%%%%%% Figure 13 Jh0Om%%%%%%%%%%%%%%%%%%%%%%%
\begin{figure}
\begin{center}
\includegraphics[width=1.0\columnwidth]{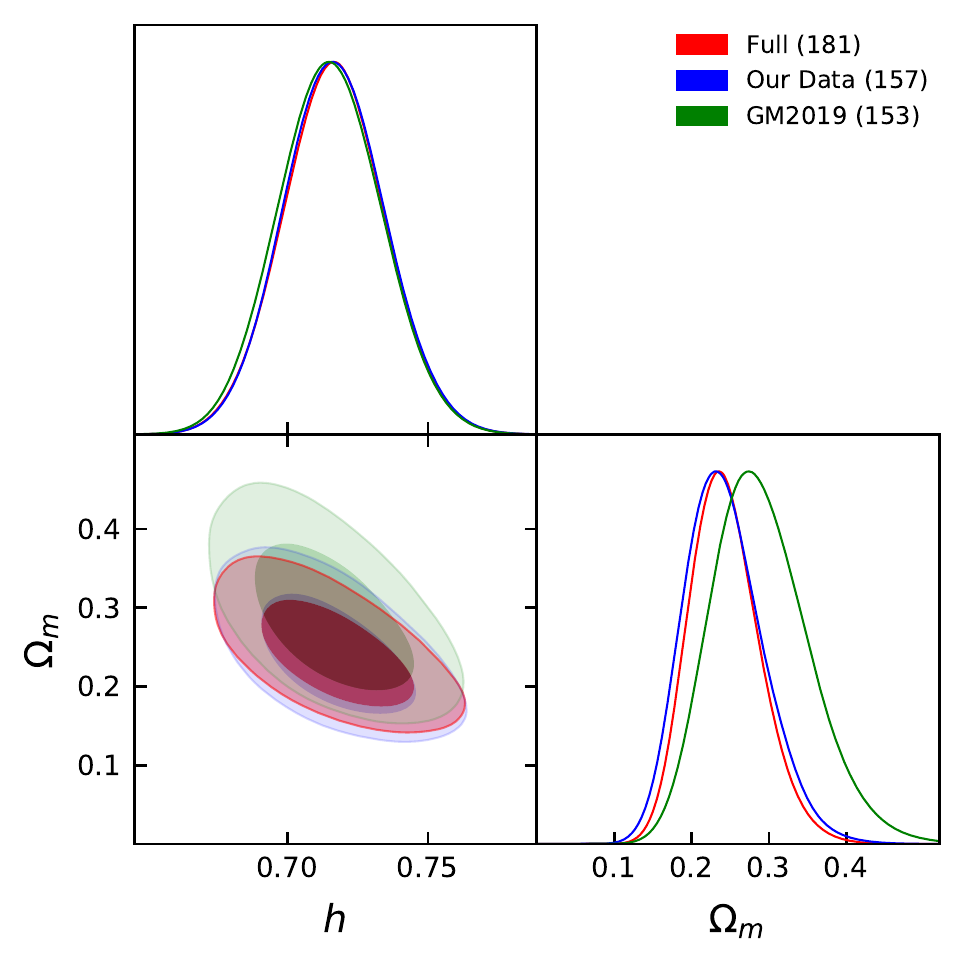}
\end{center}
        \caption{Confidence levels in the $\{h, \Omega_m\}$ plane for our current Full sample (Full), our sample excluding data from the literature (Our Data), and \citet{GonzalezMoran2019} sample (GM2019). The number of objects for each sample is shown in parentheses.}
\label{fig:Jh0Om}
\end{figure}

Using the full likelihood function as stated in Eq. \ref{eq:lkh}, it is possible to constrain the $\{h, \Omega_m \}$ plane; in this case for our Full sample we obtain:

$h=0.717 \pm 0.018$ (stat) and $\Omega_m=0.243^{+0.039}_{-0.050}$ (stat) 

\noindent
as  reported in Table \ref{tab:parLit}. 

While using Our Data sample we obtain:

$h=0.717\pm 0.018$ (stat) and $\Omega_m=0.242^{+0.042}_{-0.057}$ (stat), 

\noindent
as reported in Table \ref{tab:par}. 

If we use the 153 HIIG from \citet{GonzalezMoran2019}  to constrain the same model, we get: 

$h=0.715\pm 0.018$ (stat) and $\Omega_m=0.290^{+0.052}_{-0.070}$ (stat), 

\noindent
comparing the $FoM$ of these last results with that one from the Full sample the relative improvement of the cosmological parameters constraints is $\sim 46\%$. In Fig. \ref{fig:Jh0Om} we show a comparison for the three cases discussed above. 

Reanalysing our sample from \citet{GonzalezMoran2019}, \citet{Cao2020} also constrain the $\{h, \Omega_m \}$ plane for the $\Lambda$CDM model and they obtain:

$h=0.717 \pm 0.018$(stat) and $\Omega_m=0.289^{+0.053}_{-0.071}$ (stat), 
which is consistent with our determination for the Full sample.

\subsubsection{Constraining the dark energy equation of state}
\label{sec:ConstrainingDEEoS}
\begin{figure}
\centering
\hspace*{-0.7cm}
	\includegraphics[width=1.0
	\columnwidth]{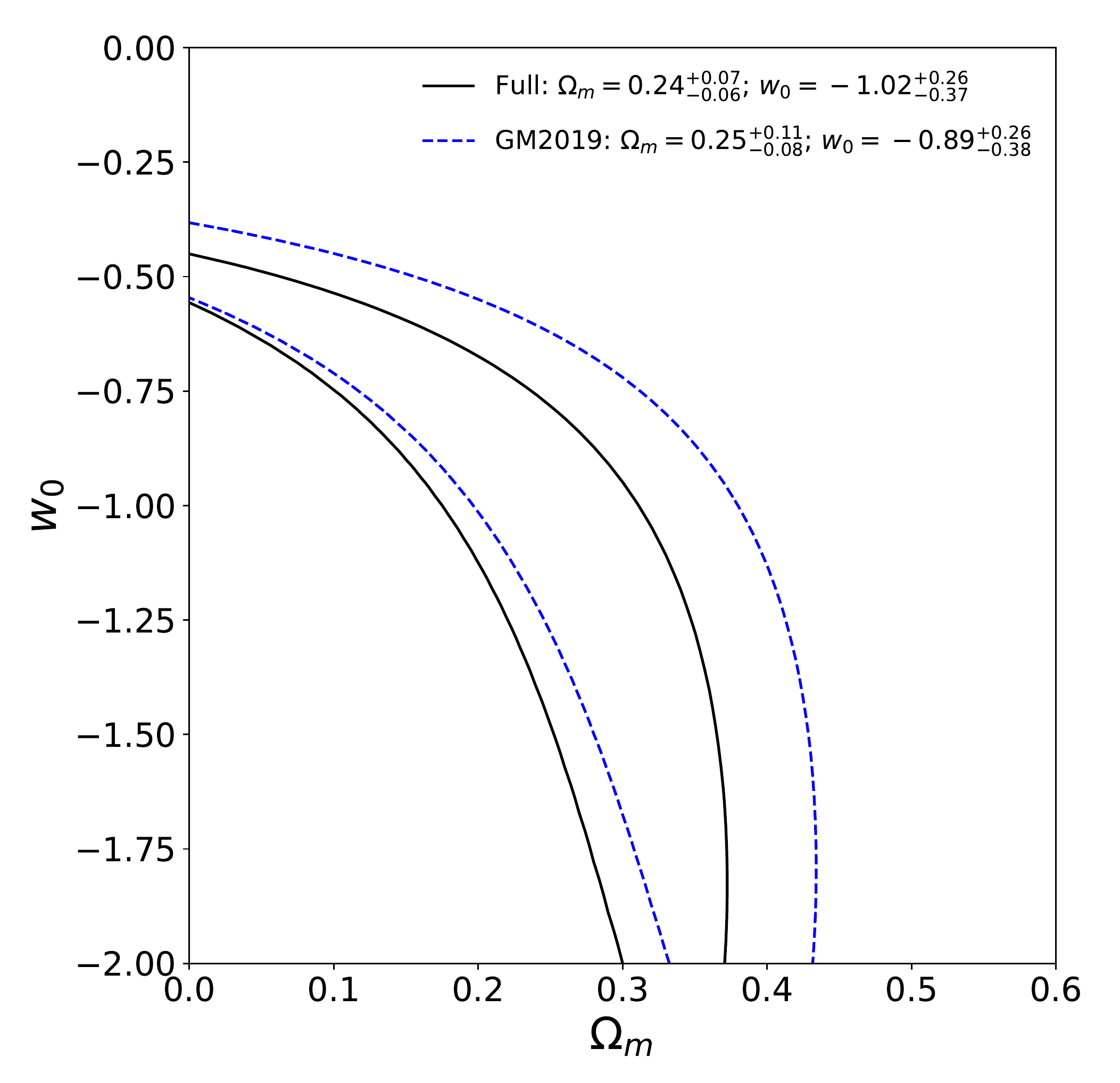}
\vspace*{-0.5cm}
\caption{$\chi^2$ likelihood contours for $\Delta\chi^2=\chi^2_{tot}-\chi^2_{tot,min}$ equal to 2.30 corresponding to the 1$\sigma$ confidence level in the $\{\Omega_m, w_0\}$ plane. The solid black line is for the Full sample and the dashed blue line is for the GM2019 one.}
\label{fig:chi2-Likelihood contours}
\end{figure}

For comparison with previous works \citep[e.g.][]{GonzalezMoran2019}, we get constraints in the $\{ \Omega_m, w_0 \}$ plane using the classical $\chi^2$-minimisation and our results are: $\Omega_m=0.24^{+0.07}_{-0.06}$ (stat) and $w_0=-1.02^{+0.26}_{-0.37}$ (stat). Fig. \ref{fig:chi2-Likelihood contours} shows the $\chi^2$ likelihood contours corresponding to the 1$\sigma$ confidence level together with that from \citet{GonzalezMoran2019} (for 181 vs. 153 objects in the same wavelength range).

%%%%%%%%%%%%%%%%%%%% Figure   14 Omw0 clasic %%%%%%%

Applying the $h$-free likelihood trough the MultiNest Bayesian sampler, as described above, to constrain the parameters of the $w$CDM model,  we  obtain the marginalised best-fit parameter values and $1\sigma$ uncertainties in the $\{\Omega_m, w_0\}$ plane for the Full sample. Our constraints are:

$\Omega_m=0.249^{+0.11}_{-0.065}$ (stat) and $w_0=-1.18^{+0.45}_{-0.41}$ (stat).

\begin{figure}
\begin{center}
\includegraphics[width=1.0\columnwidth]{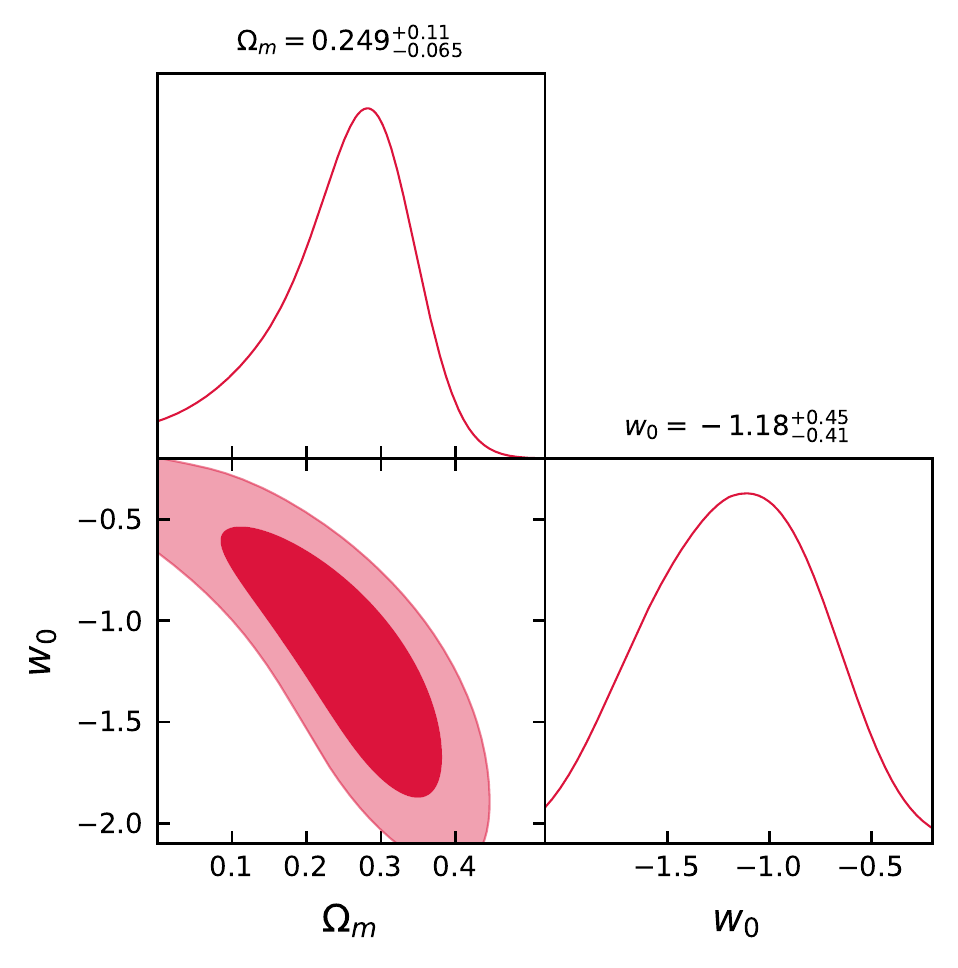}
\end{center}
\caption{Likelihood contours corresponding to 1$\sigma$ and 2$\sigma$ confidence levels in the $\{\Omega_m, w_0\}$ space for the Full sample.}
%  including data from the literature.}
\label{fig:OmW0}
\end{figure}

\noindent 
This result is presented also in Table \ref{tab:parLit} and the likelihood contours corresponding to the 1 and 2$\sigma$ confidence levels are shown in Fig. \ref{fig:OmW0}.

Following the same methodology for Our Data sample  the results are:

$\Omega_m=0.246^{+0.11}_{-0.076}$ (stat) and $w_0=-1.16^{+0.46}_{-0.40} $ (stat), 

\noindent
which are also reported in Table \ref{tab:par}. Comparing this determination with the results obtained in \citet{GonzalezMoran2019} and the $FoM$ as described above, the improvements on the cosmological parameters constraints are $\sim31\%$ and  $\sim40\%$ for the Full sample. 

%%%%%%%%%%%%%%%%%%% Figure 15 OmW0 %%%%%%%%%%%%%%%%%%%%%%%

\begin{figure}
\begin{center}
\includegraphics[width=1.0\columnwidth]{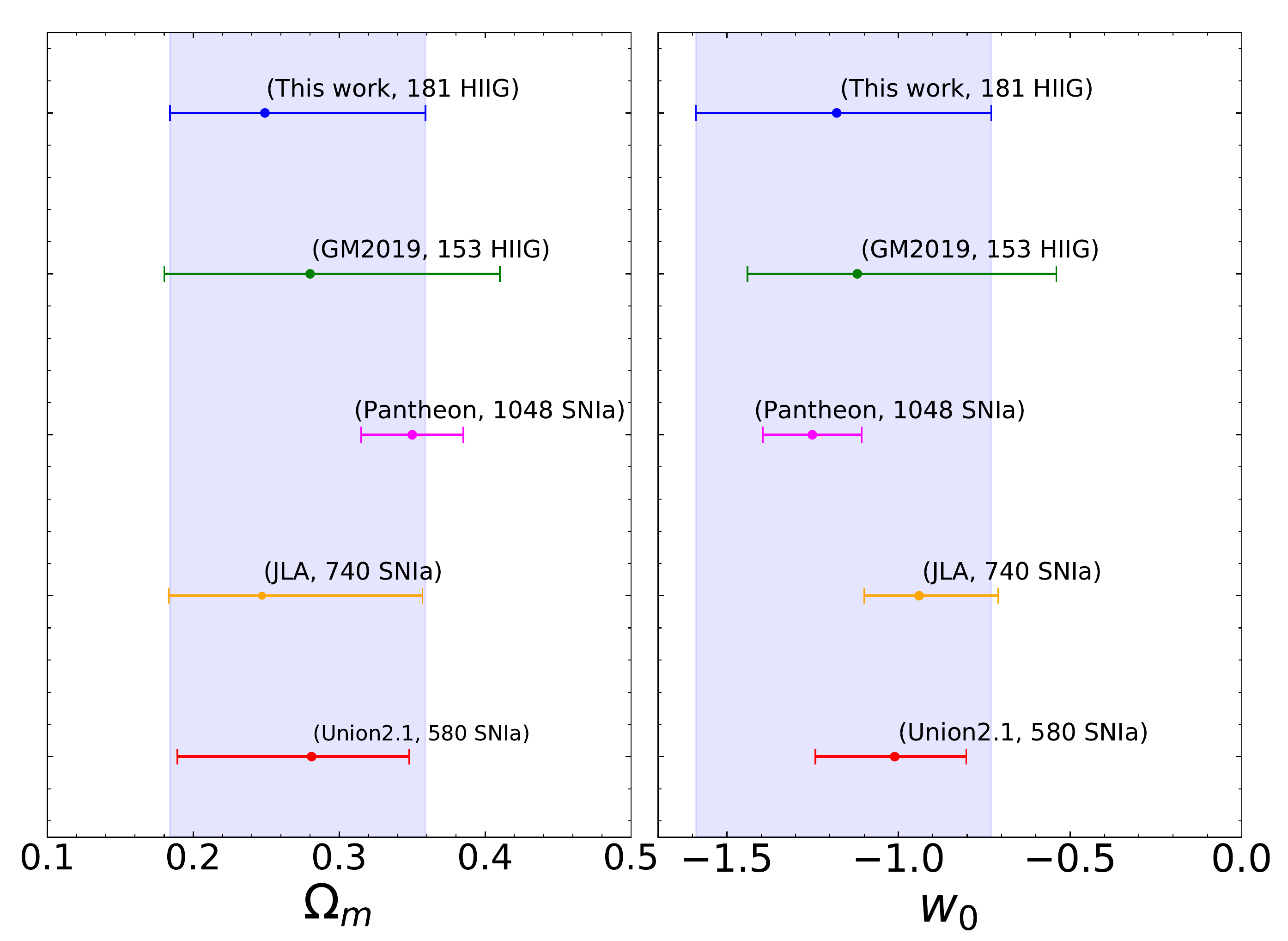}
\end{center}
        \caption{Comparison of our results for $\{\Omega_m, w_0\}$ plane (Full HIIG sample) with the literature. Only statistical uncertainties are considered. As in Fig. \ref{fig:valoresOM}, the shaded region represents the  uncertainties of our current  values.}
\label{fig:valoresOm_w0}
\end{figure}

Fig. \ref{fig:valoresOm_w0} shows our constraints for the Full sample with other recent determinations for the $\{\Omega_m, w_0\}$ plane. It is clear that our determination is fully consistent with the other results from local Universe probes, specifically SNIa. The best agreement is with the results from the JLA sample \citep{Betoule2014}, $\Omega_m=0.247^{+0.11}_{-0.064}$ (stat) and $w_0=-0.94^{+0.23}_{-0.16}$ (stat), and with the Union2.1 sample \citep{Suzuki2012}, $\Omega_m=0.281^{+0.067}_{-0.092}$ (stat) and $w_0=-1.011^{+0.208}_{-0.231}$ (stat). The results from the most recent Pantheon sample \citep{Scolnic2018}, $\Omega_m=0.350\pm0.035$ (stat) and $w_0=-1.251\pm0.144$ (stat), produce a considerably larger value for $\Omega_m$ than both previous SNIa samples and our own  determination. Even so, we are still marginally consistent due partially to our large error bars. One point of interest is the big drift on the values for the $\{\Omega_m, w_0\}$ plane from the JLA to the Pantheon SNIa samples, which may be interesting to explore.

%%%%%%%%%%%%%%%%%%% Figure 16 Om-w0 comparison %%%%%%%%%%%%%%%%%%%%%%%

\begin{figure}
\begin{center}
\includegraphics[width=1.0\columnwidth]{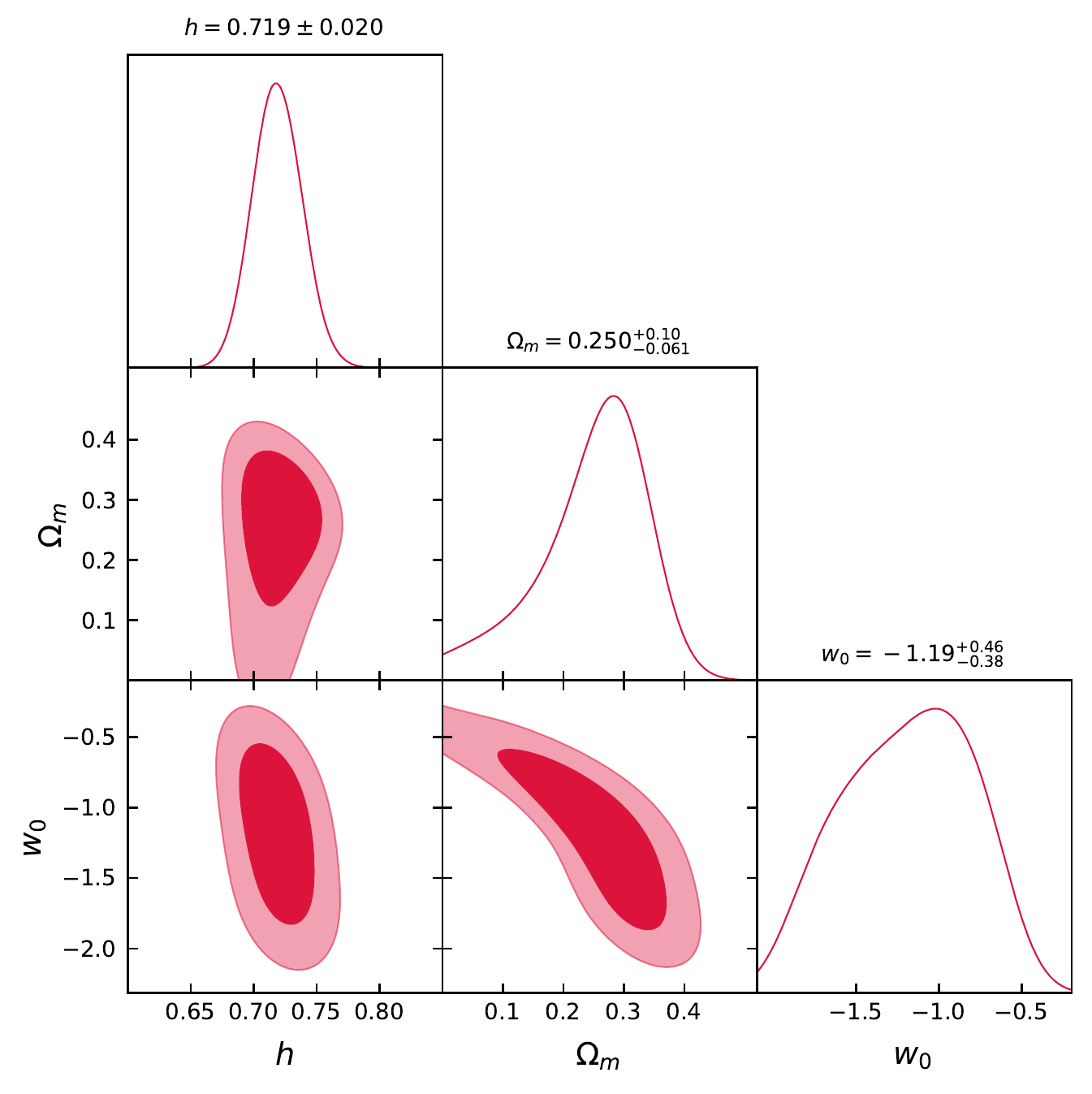}
\end{center}
\caption{Likelihood contours corresponding to the 1$\sigma$ and 2$\sigma$ confidence levels in the $\{h, \Omega_m, w_0\}$ space for the Full sample.}
\label{fig:h0Omw0}
\end{figure}

As in the previous section, we also analyse the results from the full likelihood to constrain the $\{h, \Omega_m, w_0\}$ plane. When we use the Full sample, our determination is $\{h, \Omega_m, w_0\} = \{0.719\pm 0.020, 0.250^{+0.10}_{-0.061}, -1.19^{+0.46}_{-0.38} \}$ (stat)  (Fig. \ref{fig:h0Omw0} and in Table \ref{tab:parLit}). Using Our Data sample, the result is $\{h, \Omega_m, w_0\} = \{0.718\pm 0.020, 0.245^{+0.11}_{-0.071}, -1.16^{+0.50}_{-0.35} \}$ (stat).

%%%%%%%%%%%%%%%%%%% Figure 17 h0Omw0 %%%%%%%%%%%%%%%%%%%%%%%

\subsubsection{Nuisance and cosmological parameters simultaneous determination}
\label{sec:HIIG and GHIIR}
A global fit of all the free parameters, nuisance and cosmological, applying the full likelihood to our Global sample (Table \ref{tab:Samples used in the HIIG cosmological analysis}) provides the following results: \\

\begin{figure*}
\begin{center}
\includegraphics[width=1.5\columnwidth]{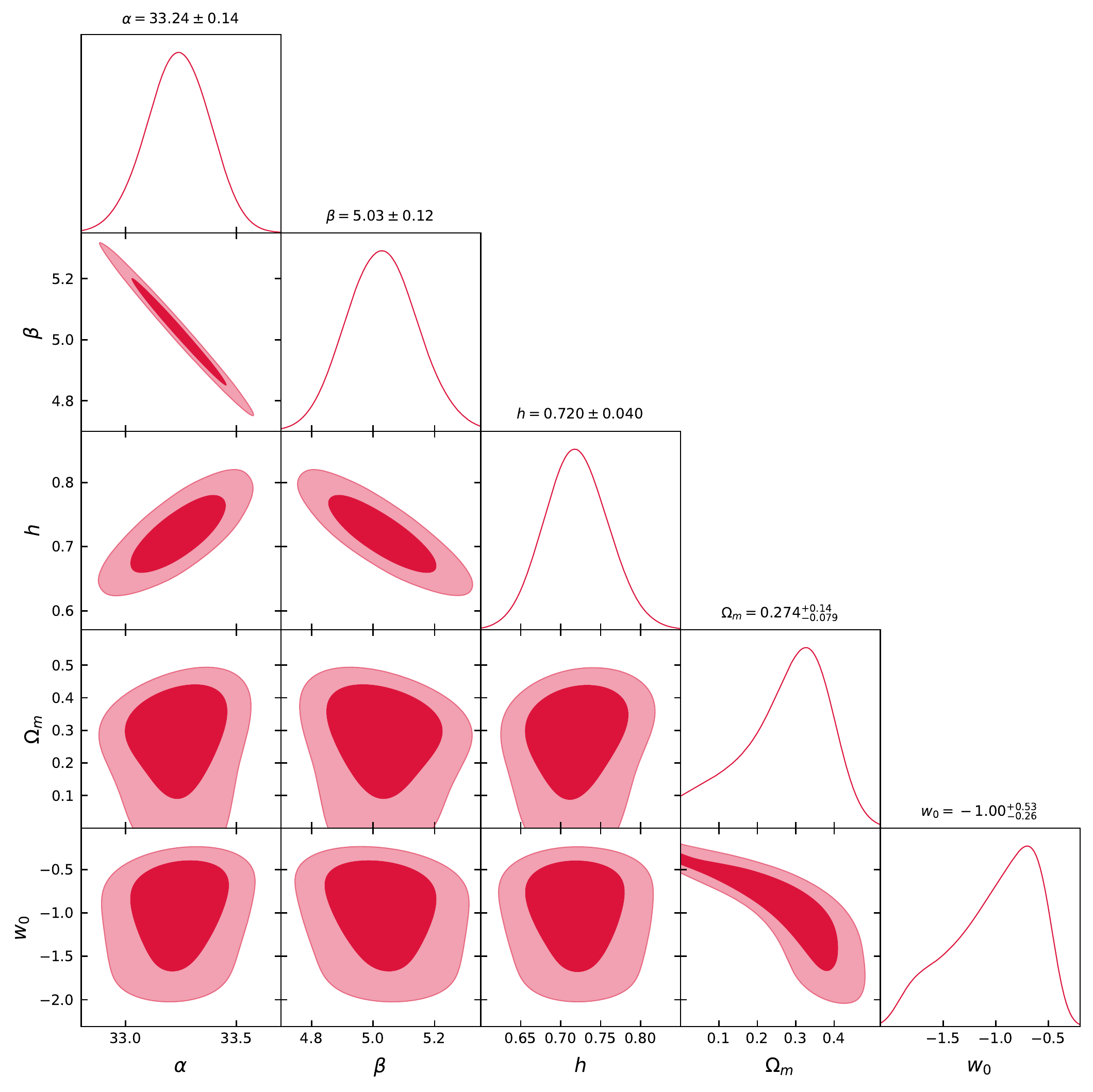}
\end{center}
\caption{Likelihood contours corresponding to the 1$\sigma$ and 2$\sigma$ confidence levels in the $\{\alpha, \beta, h,\Omega_m\, w_0\}$ space for the Global sample.}
\label{fig:abhOmW0}
\end{figure*}

$\alpha$ = 33.24 $\pm$ 0.14, $\beta$ = 5.03 $\pm$ 0.12, 

$h$ = 0.720 $\pm$ 0.040, 

$\Omega_m = 0.274^{+0.14}_{-0.079}$ , $w_0 = -1.00^{+0.53}_{-0.26}$, \\

\noindent
as reported in Table \ref{tab:parLit}. In Fig. \ref{fig:abhOmW0} we plot the 1 and 2$\sigma$ likelihood contours in the various parameter planes. From this result, it is clear that our determination of the global set of parameters is fully consistent with the previous determinations cited above and with other recent determinations. Comparing the $FoM$ of these results with those obtained in \citet{GonzalezMoran2019} the improvement is $\sim52\%$. 

Following the same methodology, we also present in Table \ref{tab:par} the results of excluding the Literature sample from the Global sample, i.e., for 193 objects.

%%%%%%%%%%%%%%%%%%% Figure 18 abhOmW0 %%%%%%%%%%%%%%%%%%%%%%%

%%%%%%%%%%%%%%%% Table 5 Results Full  %%%%%%%%%%%%%%%%%%
\begin{table*}
        \caption{Marginalised best-fit parameter values and $1\sigma$ uncertainties for the HIIG sample, including data taken from the literature, and its combination with other samples. In parenthesis the fixed values.}
\label{tab:parLit}
\resizebox{\textwidth}{!}{
\begin{tabular}{lccccccccc}
\hline
\hline
	Data Set       & $\alpha$ & $\beta$ & $\omega_b$ & $h$ & $\Omega_m$ & $w_0$ & $w_a$  & N & $\chi^2$ \\ %& $FoM$ \\
\hline
	HIIG & --- & ($5.022\pm 0.058$) & --- & --- & $0.244^{+0.040}_{-0.049}$ & (-1.0) & (0.0) & 181 & 258.11 \\ % & 20.82 \\
	HIIG & --- & ($5.022\pm 0.058$) & --- & --- & $0.249^{+0.11}_{-0.065}$ & $-1.18^{+0.45}_{-0.41}$ & (0.0) & 181 & 258.33 \\ % & 43.5 \\
	HIIG & ($33.268\pm 0.083$) & ($5.022\pm 0.058$) & --- & $0.717\pm 0.018$ & $0.243^{+0.039}_{-0.050}$ & (-1.0) & (0.0) & 181 & 258.10 \\% & 1459.87 \\
	HIIG & ($33.268\pm 0.083$) & ($5.022\pm 0.058$) & --- & $0.719\pm 0.020$ & $0.250^{+0.10}_{-0.061}$ & $-1.19^{+0.46}_{-0.38}$ & (0.0) & 181 & 258.55 \\% & 2714.68 \\
	HIIG & $33.23\pm 0.14$ & $5.04\pm 0.12$ & --- & $0.719\pm 0.040$ & $0.313^{+0.046}_{-0.057}$ & (-1.0) & (0.0) & 217 & 785.12\\% & 862561.4 \\
	HIIG & $33.24\pm 0.14$ & $5.03\pm 0.12$ & --- & $0.720\pm 0.040$ & $0.274^{+0.14}_{-0.079}$ & $-1.00^{+0.53}_{-0.26}$ & (0.0) & 217 & 786.41\\% & 1232154.05 \\
\hline
	SNIa & --- & --- & --- & --- & $0.349^{+0.038}_{-0.029}$ & $-1.25^{+0.15}_{-0.13}$ & (0.0) & 1048 & 1032.56\\% & 552.17 \\
\hline
	BAO  & --- & --- & (0.02225) & (0.6774) & $0.306^{+0.035}_{-0.041}$ & $-0.96^{+0.11}_{-0.14}$ & (0.0) & 6 & 1.76\\% & 364.30 \\
\hline
	CMB  & --- & --- & $0.02225\pm 0.00016$ & (0.6774) & $0.3107\pm 0.0029$ & $-0.954\pm 0.020$ & (0.0) & 3 & 0.025\\% & 54256.65 \\
\hline
	HIIG+CMB+BAO & --- & ($5.022\pm 0.058$) & (0.02225) & $0.693\pm 0.01$ & $0.298\pm 0.012$ & $-1.005\pm 0.051$ & (0.0) & 190 & 262.37\\% & 6225320.03 \\
	HIIG+CMB+BAO & --- & ($5.022\pm 0.058$) & (0.02225) & $0.698\pm 0.023$ & $0.294^{+0.018}_{-0.021}$ & $-1.07^{+0.23}_{-0.32}$ & $0.16^{+0.96}_{-0.55}$ & 190 & 262.14\\% & 2358318.12 \\
\hline
	SNIa+CMB+BAO & --- & --- & (0.02225) & $0.6977\pm 0.0060$ & $0.2940\pm 0.0057$ & $-1.024\pm 0.022$ & (0.0) & 1057 & 1037.80\\% & 22743679.31 \\
	SNIa+CMB+BAO & --- & --- & (0.02225) & $0.6885\pm 0.0099$ & $0.3011\pm 0.0085$ & $-1.062\pm 0.040$ & $0.25^{+0.22}_{-0.19}$ & 1057 & 1036.37\\% & 82548459.35 \\
\hline 
\hline
\end{tabular}}
\end{table*}

%%%%%%%%%%%%%%%% Table 4 Results No Literature %%%%%%%%%%%%%%%%%%
\begin{table*}
        \caption{Marginalised best-fit parameter values and $1\sigma$ uncertainties for our own HIIG sample. Fixed values  in parenthesis.}
\label{tab:par}
\resizebox{\textwidth}{!}{
\begin{tabular}{lcccccccc}
\hline
\hline
	Data Set       & $\alpha$ & $\beta$ & $h$ & $\Omega_m$ & $w_0$ & $w_a$  & N & $\chi^2$ \\ %& $FoM$ \\
\hline
	HIIG & --- & ($5.022\pm 0.058$) & --- & $0.243^{+0.047}_{-0.057}$ & (-1.0) & (0.0) & 157 & 231.09 \\%& 18.20 \\
	HIIG & --- & ($5.022\pm 0.058$) & --- & $0.246^{+0.11}_{-0.076}$ & $-1.16^{+0.46}_{-0.40}$ & (0.0) & 157 & 231.25 \\%& 39.4 \\
	HIIG & ($33.268\pm 0.083$) & ($5.022\pm 0.058$) & $0.717\pm 0.018$ & $0.242^{+0.042}_{-0.057}$ & (-1.0) & (0.0) & 157 & 231.10 \\%& 1224.80 \\
	HIIG & ($33.268\pm 0.083$) & ($5.022\pm 0.058$) & $0.718\pm 0.020$ & $0.245^{+0.11}_{-0.071}$ & $-1.16^{+0.50}_{-0.35}$ & (0.0) & 157 & 231.43 \\%& 2411.53 \\
	HIIG & $33.25\pm 0.14$ & $5.02\pm 0.12$ & $0.723\pm 0.041$ & $0.322^{+0.049}_{-0.061}$ & (-1.0) & (0.0) & 193 & 733.41 \\%& 746686.10 \\
	HIIG & $33.25\pm 0.14$ & $5.02\pm 0.12$ & $0.723\pm 0.040$ & $0.276^{+0.14}_{-0.091}$ & $-0.98^{+0.55}_{-0.21}$ & (0.0) & 193 & 734.42 \\%& 1107398.92 \\
\hline
\hline
\end{tabular}}
\end{table*}

\subsection{Systematic errors}
\label{sec:Discussion}

In previous sections we have discussed the statistical uncertainties associated with our methodology. However, the scatter found in the \lsigG\ relation for HIIG suggests the presence of a second parameter probably associated with the line profile shape \citep{Bordalo_Telles2011,Chavez2014}
or/and systematic errors. Systematic uncertainties are difficult to estimate and in this section we will briefly discuss part of the systematic errors that can be included in the likelihood function (Eq. \ref{eq:epsilon}).

Not knowing the shape of the extinction law for HIIG and its possible variation with redshift is an important source of uncertainty.  We have found \citet{GonzalezMoran2019} that when applying to the data a correction based on the \citet{Calzetti2000}  law, we obtain a smaller reduced $\chi^2$ (1.1) than when we apply the \citet{Gordon2003} correction (1.7). However, Calzetti's law was derived from a sample of eight heterogeneous starburst galaxies where only two, Tol 1924-416 and UGCS410 are bonafide HIIG and the rest are evolved high metallicity starburst galaxies, while  \citet{Gordon2003} extinction curve corresponds to the LMC supershell near the 30 Doradus star forming region, the prototypical GHIIR. Besides, as already mentioned in \S \ref{sec:Extinction correction}, the dust attenuation curve derived from analogues of high-redshift star-forming galaxies agrees quite well with \citet{Gordon2003}. Therefore, we prefer the results using  Gordon's extinction curve. 
 
A related source of uncertainty is associated with the fact that we do not have extinction estimates for individual HIIG with $z \geq 1$ and for these systems we have adopted the average extinction of the low-z HIIG assuming that there is no systematic variation associated with redshift. A welcome improvement would be to obtain the Balmer decrement for the high-z sample. 
 
In \citet{Chavez2016} we presented a systematic error budget on the distance moduli of 0.257. This includes the typical uncertainty contribution from the size and age of the burst, abundances and extinction. Adding in quadrature this systematic error budget in Eq. \ref{eq:epsilon}, we obtain a reduced $\chi^2$ close to 1 and an estimate of a systematic uncertainty of 0.02 in the $\Omega_m$ parameter.

The origin of the small difference between Balmer and [$\Oiii$]$\lambda$5007\AA\ lines velocity dispersion (presented in \ref{velocity dispersion}) is still unknown \citep[e.g.][]{Hippelein1986, Bordalo_Telles2011, Bresolin2020} and induces a systematic error that needs to be analysed. Constraining only the $\Omega_m$ parameter and the $\{\Omega_m, w_0\}$ plane without applying the transformation $\sigma$([OIII])/$\sigma$(H$\beta$) yields $\Omega_m=0.256^{+0.042}_{-0.052}$ and $\{\Omega_m, w_0\}=\{0.258^{+0.11}_{-0.066},-1.17^{+0.46}_{-0.41}\}$, respectively. Comparing these values with those given in Table \ref{tab:parLit}, the results are in agreement at better than 1$\sigma$ level and give us an estimate of a systematic uncertainty of 0.01 in both the $\Omega_m$ parameter and the $\{\Omega_m, w_0\}$ plane.

A full discussion of the complex analysis of systematic errors in the \lsigG\ method will be the subject of a forthcoming paper (Ch\'avez et al. in prep.).

\subsection{Joint analysis}
\label{sec:Joint analysis}
A joint-likelihood analysis with the CMB and 
BAO probes is performed on the Full sample using the $h$-free likelihood method. We also compare our results with the combination of SNIa, CMB and BAO \citep[as in][]{GonzalezMoran2019} except that we use here  the Pantheon sample instead of the JLA one.

\begin{figure*}
\begin{center}$
\begin{array}{cc}
  \subfloat[HIIG + CMB + BAO] {\includegraphics[width=.5\textwidth]{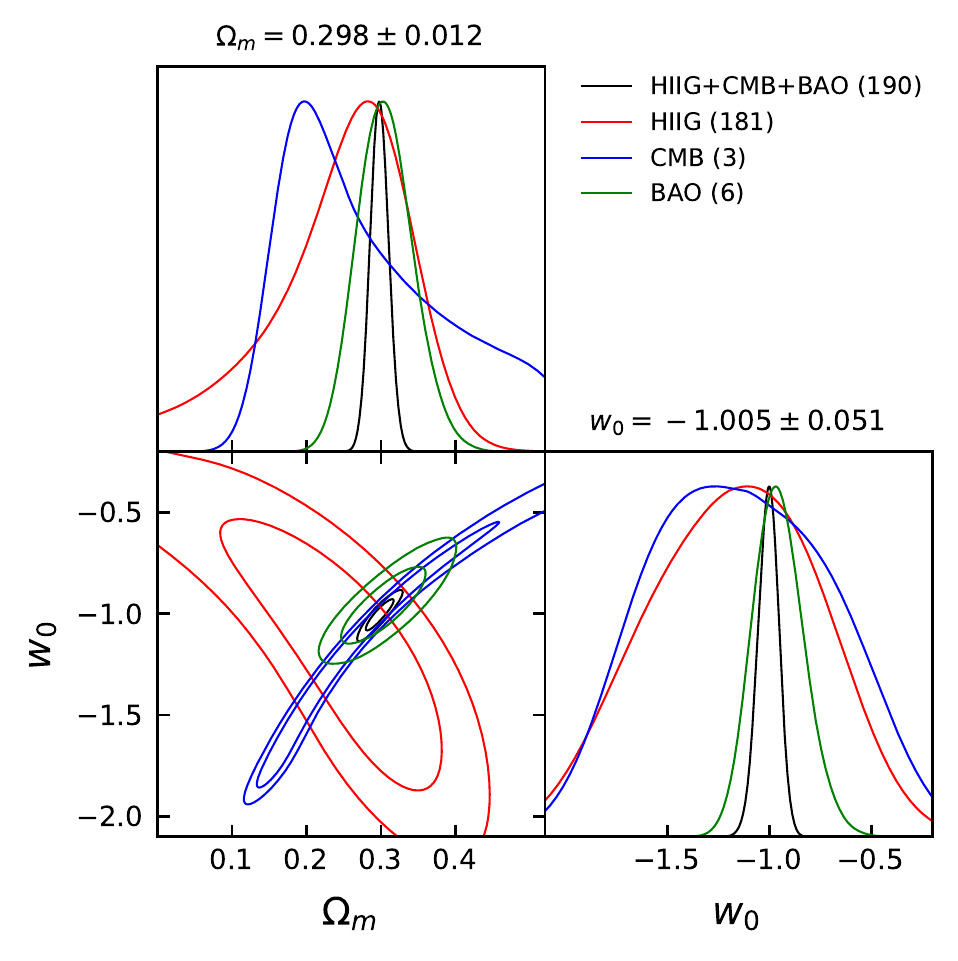}} &
  \subfloat[SNIa + CMB + BAO] {\includegraphics[width=.5\textwidth]{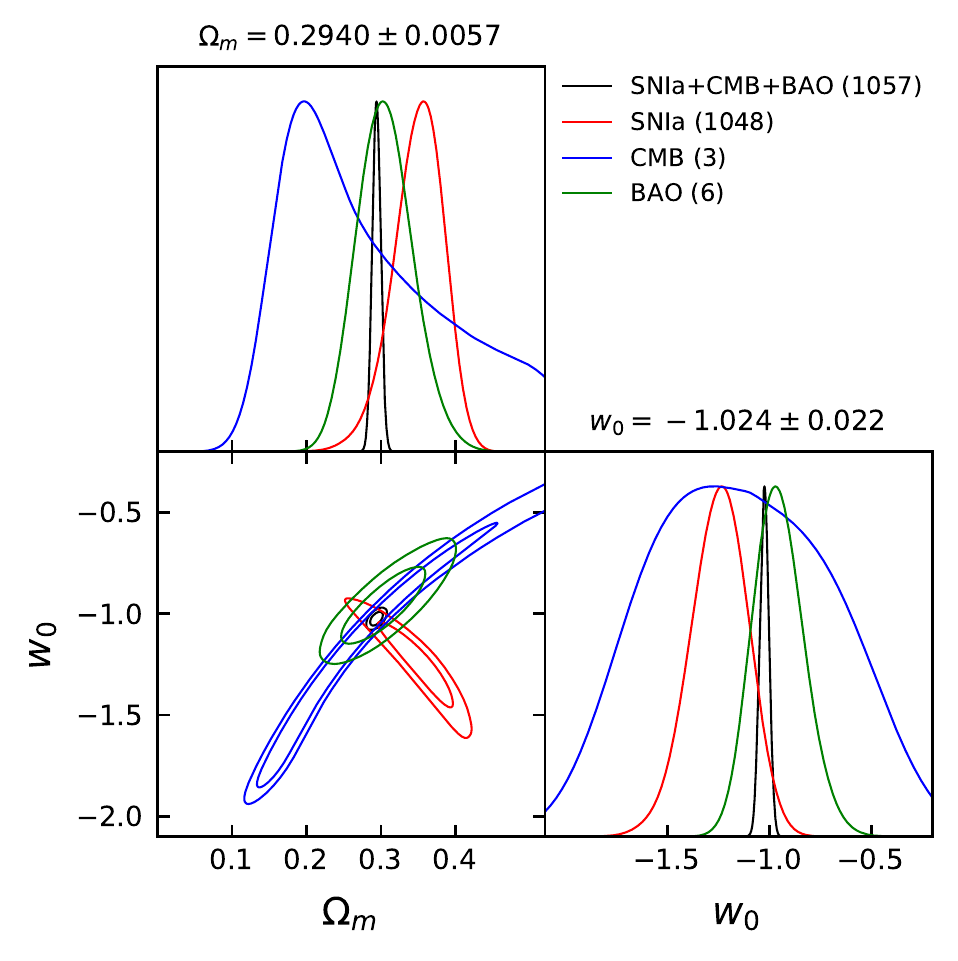}}
\end{array}$
\end{center}
\caption{Likelihood contours corresponding to the 1$\sigma$ and 2$\sigma$ confidence levels in the $\{\Omega_m, w_0\}$ space for a) the joint sample of HIIG, CMB and BAO and b) the joint sample of SNIa, CMB and BAO. We show in the inset the sample size used in the analysis. Only statistical uncertainties are shown.}
\label{fig:JOmW0}
\end{figure*}

The joint analysis in the $\{\Omega_m, w_0\}$ plane is shown in Fig.  \ref{fig:JOmW0} for HIIG, CMB and BAO in  panel (a)  and for  SNIa, CMB and BAO in  panel (b). The figure shows similar inclinations for HIIG and SNIa, perhaps due to the fact 
that both distance estimators restrict the solutions space in a comparable redshift range.
 
Combining HIIG, CMB and BAO yields: \\

$\Omega_m=0.298 \pm 0.012$ and $w_0=-1.005 \pm 0.051$,\\

\noindent
fully consistent with the $\Lambda$CDM model.
From Fig.  \ref{fig:JOmW0} and Table \ref{tab:parLit} it is clear that the solution space of HIIG/CMB/BAO, although less constrained, is certainly compatible with the solution space of SNIa/CMB/BAO.

We have explored the possibility of constraining the evolution of the DE with time for the CPL parametrisations, using a joint analysis of the HIIG, CMB and BAO measurements, which leaves the relevant parameters mostly unconstrained. However if we marginalise one over the other, we obtain:\\

\begin{figure*}
\begin{center}$
\begin{array}{cc}
  \subfloat[wCDM] {\includegraphics[width=.5\textwidth]{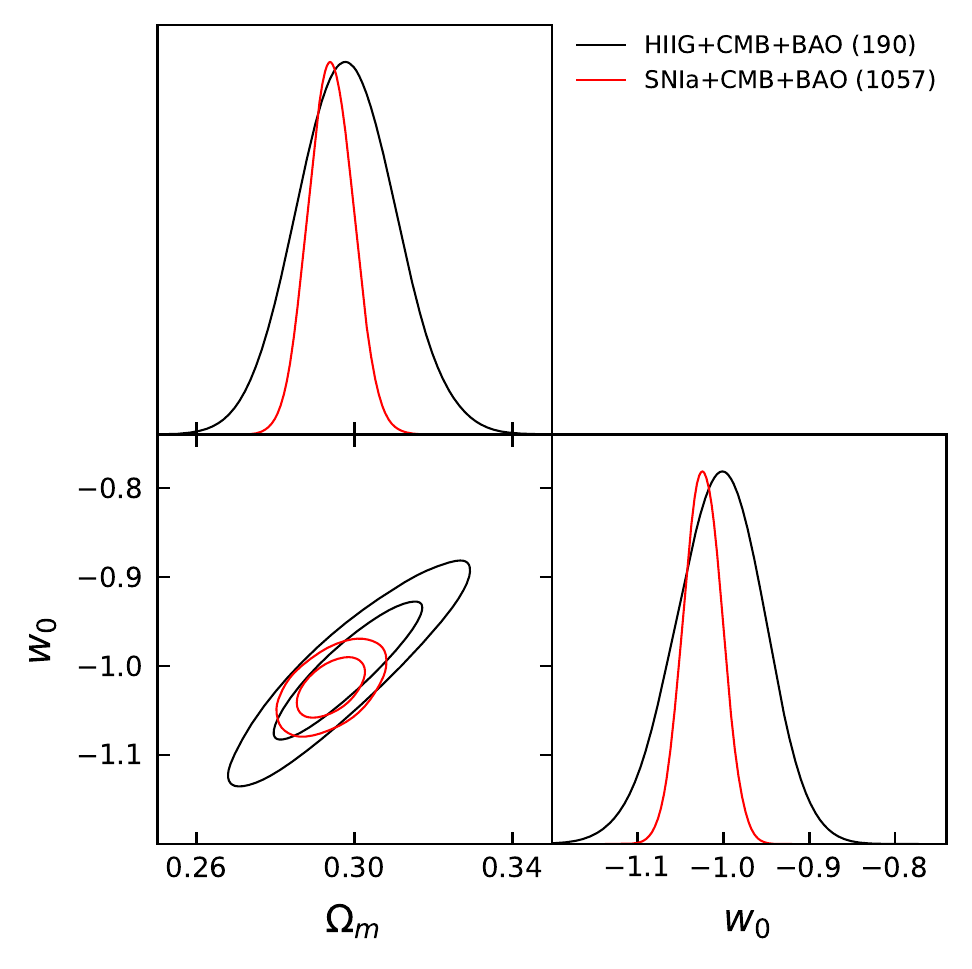}} &
  \subfloat[CPL] {\includegraphics[width=.5\textwidth]{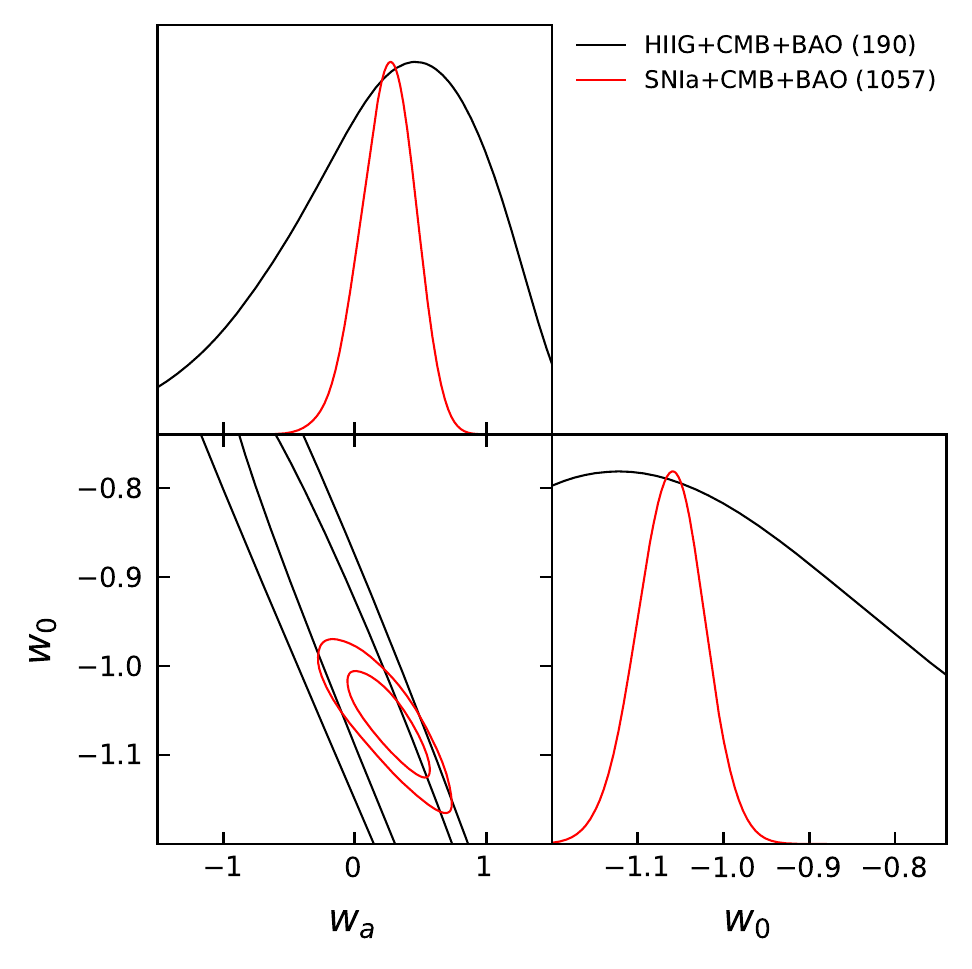}}
\end{array}$
\end{center}
        \caption{Joint likelihood contours of  HIIG/CMB/BAO (black contours) and  SNIa/CMB/BAO (red contours) probes. (a): wCDM DE EoS parametrisation (\{$h$, $\Omega_m$, $w_0$\}). (b): CPL DE EoS parametrisation (\{$h$, $\Omega_m$, $w_0$, $w_a$\}).}
\label{fig:j2}
\end{figure*}

$w_0=-1.07^{+0.23}_{-0.32}$, $w_a = 0.16^{+0.96}_{-0.55}$, \\

\noindent
which --although with large uncertainties-- are consistent with no evolution.
The joint likelihood contours for the probes HIIG/BAO/CMB and SNIa/BAO/CMB are shown in Fig. \ref{fig:j2} and reported in Table \ref{tab:parLit} for wCDM and CPL DE EoS parameterisations. It is clear that the HIIG/BAO/CMB and SNIa/BAO/CMB joint probes agree with each other for both parameterisations although the latter produces better constraints, which is to be expected given the much larger number of SNIa (1048) than HIIG (181) used.

In a forthcoming paper (Tsiapi et al. in prep.), we will compare the HIIG results against the full CMB spectrum as provided by \citet{PlanckCollaboration2018}, in order to place constraints on the whole set of cosmological parameters including those of the angular size of the sound horizon at recombination ($\theta_{MC}$), the amplitude of the primordial power spectrum ($A_s$), the spectral index ($n_s$) and the optical depth at reionisation ($\tau$).

%%%%%%%%%%%%%%%%%% Figure 19 JointOmW0 %%%%%%%%%%%%%

%%%%%%%%%%%%%%%%%%%% Figure  20 Joint OmW0, W0Wa %%%%%%%%%%%%%%%

\section{Conclusions}
\label{sec:Conclusions}

We have analysed a set of 181 HIIG in the redshift range 0.01 $<$ z $<$ 2.6. The sample includes a new set of 41 HIIG observed with KMOS at the VLT in the range of redshift 1.3 $<$ z $<$ 2.6. Using the L - $\sigma$ distance indicator we have constrained cosmological parameters independently of the value of the Hubble constant.

Regarding the restrictions of the $\Omega_{m}$ parameter
% independently of the value of the Hubble constant, 
we found that HIIG alone constrain the matter density to high significance. Using the Full sample of 181 HIIG and the $\chi^2$-minimisation procedure, we find:\\

$\Omega_m = 0.236^{+0.047}_{-0.041}$ (stat),\\

\noindent
while using the MultiNest MCMC procedure we find:\\

$\Omega_m = 0.244^{+0.040}_{-0.049}$ (stat),\\

\noindent
an improvement of the cosmological parameter constraints by 37\% with respect to the value published in \citet{GonzalezMoran2019} for a sample of 153 HIIG.

HIIG also constrain the value of the DE EoS parameter in the $\{\Omega_m, w_0\}$ plane independently of the value of the Hubble constant. 
The marginalised best-fit values for the Full sample using the $\chi^2$-minimisation procedure are: \\

$\Omega_m=0.24^{+0.07}_{-0.06}$ and $w_0=-1.02^{+0.26}_{-0.37}$ (stat), \\

\noindent
and using the MultiNest MCMC procedure are:\\

$\Omega_m=0.249^{+0.11}_{-0.065}$ and $w_0=-1.18^{+0.45}_{-0.41}$ (stat), \\

\noindent
an improvement of the cosmological parameters constraints by 40\% over our previous results in \citet{GonzalezMoran2019}. Our present results are in accordance with those based on the JLA SNIa sample \citep{Betoule2014} of $\Omega_m=0.247^{+0.11}_{-0.064}$ and $w_0=-0.94^{+0.23}_{-0.16}$ (stat) at a $<1\sigma$ level.

Even if our main objective was to constrain the DE EoS parameter jointly with the $\Omega_m$ parameter independently of the Hubble constant, prompted by the tension on estimates of the latter, we used the Full sample to constrain the parameter space in the \{$h,\Omega_m$\} plane. The resulting preferred values in the $\Lambda$CDM scenario are:\\

 $h=0.717 \pm 0.018$ and $\Omega_m=0.243^{+0.039}_{-0.050}$ (stat). \\
 
 \noindent
 These values favour the late Universe results given by SNIa over the early Universe ones given by the analysis of Planck-CMB data.

Using simultaneously GHIIR and HIIG data (217 objects) a global fit of all the free parameters, nuisance and cosmological, provides: \\

$\alpha$ = 33.24 $\pm$ 0.14, $\beta$ = 5.03 $\pm$ 0.12, 

$h$ = 0.720 $\pm$ 0.040, 

$\Omega_m = 0.274^{+0.14}_{-0.079}$ , $w_0 = -1.00^{+0.53}_{-0.26}$, \\

 \noindent
in agreement with the Union 2.1 sample results \citep{Suzuki2012} of $\Omega_m=0.281^{+0.067}_{-0.092}$ and $w_0=-1.011^{+0.208}_{-0.231}$ (stat).

Combining HIIG, CMB and BAO yields our best estimates: \\

$\Omega_m=0.298 \pm 0.012$ and $w_0=-1.005 \pm 0.051$, \\

\noindent
which, although less constrained, are certainly compatible with the solution space of SNIa/CMB/BAO using the SNIa Pantheon sample \citep{Scolnic2018}.

What is highly encouraging is the consistency of our results with those from SNIa, CMB and BAO, particularly considering that the analysis of less than 200 HIIG was involved.
The HIIG results are comparable with those obtained from SNIa a decade ago when the sample of SNIa was around few hundred. This is consistent with the main conclusion of \citet{Plionis2011} that a sample of at least 500 HIIG (which we aim to procure in the forthcoming step of the project) is needed to have errors comparable with those of the SNIa approach. 

\section*{DATA AVAILABILITY}
The data underlying this article are available in the article and in its online supplementary material. Other datasets were derived from sources in the public domain, \cite {Chavez2014} at https://doi.org/10.1093/mnras/stu987, \cite{Terlevich2015} at  https://doi.org/10.1093/mnras/stv1128, \cite{Fernandez2018} at https://doi.org/10.1093/mnras/stx2710 and \cite{GonzalezMoran2019} at https://doi.org/10.1093/mnras/stz1577.

\section*{Acknowledgements}
ALGM is grateful to the Mexican Research Council (CONACYT) for supporting this research under studentship 419392 and to  SNI-CONACYT  in the form of a graduate assistantship. ALGM thanks Michael Hilker, ESO User Support Astronomer, for his help during the process of reducing the KMOS data. Part of this work was done under the umbrella of the Guillermo Haro Program of Advanced Astrophysics at INAOE. We are thankful to an anonymous referee for stimulating suggestions during the revision process.

\bibliography{bib/bibpaper2019}
\label{lastpage}

\appendix

 \section{Results for individual objects}
This appendix presents the new KMOS data, the parameter measurements and the total sample used for the cosmological analysis.

%%%%%%%%%%%%%%%%%%% Table1 %%%%%%%%%%%%%%%%%%%%%%%

\begin{table*}
	\small
	\centering
	\caption{KMOS  observed sample.}
	\label{tab:table1}
	\begin{tabular}{lcccccc} 
	\hline
Name	 & R.A. & DEC. & Field & Seeing$^{b}$ & Exp. time &  Candidate \\
& (J2000) & (J2000) & & (arcsec) &(seconds)&source \\
		\hline
Q2343-BM133	&	23 46 16.18	&	$+$12 48 09.31	&	Q2343	&	0.74	&	9,600	&	1,2	\\
Q2343-BM181	&	23 46 27.03	&	$+$12 49 19.65	&		&	0.86	&	19,200	&	1,2	\\
Q2343-BX163$^a$	&	23 46 04.78	&	$+$12 45 37.78	&		&	0.74	&	1,200	&	1,2	\\
Q2343-BX182	&	23 46 18.04	&	$+$12 45 51.11	&		&	0.74	&	9,600	&	1,2	\\
Q2343-BX236	&	23 46 18.71	&	$+$12 46 15.97	&		&	0.74	&	9,600	&	1,2	\\
Q2343-BX336	&	23 46 29.53	&	$+$12 47 04.76	&		&	0.98	&	9,600	&	1,2	\\
Q2343-BX341	&	23 46 23.24	&	$+$12 47 07.97	&		&	0.86	&	19,200	&	1,2	\\
Q2343-BX378$^a$	&	23 46 33.90	&	$+$12 47 26.20	&		&	0.98	&	1,200	&	1,2	\\
Q2343-BX389	&	23 46 28.90	&	$+$12 47 33.55	&		&	0.98	&	9,600	&	1,2	\\
Q2343-BX390	&	23 46 24.72	&	$+$12 47 33.80	&		&	0.98	&	9,600	&	1,2	\\
		\hline
		\multicolumn{7}{l}{}\\
		\multicolumn{7}{l}{$^a$ No emission lines detected in the KMOS data cube.}\\
		\multicolumn{7}{l}{$^b$ Seeing in the V-band corrected by airmass, as derived from the image headers.}\\
		\multicolumn{7}{l}{References: (1): \cite{Erb2006a}; (2): \cite{Erb2006b}, (3): \cite{Maseda2013};}\\
		\multicolumn{7}{l}{(4): \cite{Maseda2014}; (5): \cite{van_der_Wel2011}; (6): \cite{Mancini2011};}\\
		\multicolumn{7}{l}{(7): \cite{ForsterSchreiber2009}; (8): \cite{Xia2012}}.\\
		\multicolumn{7}{l}{The full version of this table is available as supplementary material.}\\
		\multicolumn{7}{l}{}\\
	\end{tabular}
\end{table*}

%%%%%%%%%%%%%%%%%%% Table 2 %%%%%%%%%%%%%%%%%%%%%%%

\begin{table*}
	\small
	\centering
	\caption{Measurements for the KMOS sample.}
	\label{tab:Observational data}
	\begin{tabular}{lcccccccc} 
	\hline \hline
Target &	redshift 	&	$\sigma$(H$\alpha)^a_{obs}$	&	$\sigma[\Oiii]^a_{obs}$	 &	$\sigma$(H$\beta)^a_{obs}$	&	$\sigma_{inst}$	&	F(H$\alpha)^b_{obs}$	&	F[$\Oiii]^b_{obs}$ 	&	F(H$\beta)^b_{obs}$ \\
& & [km/s] & [km/s] & [km/s] & [km/s] &  &  &  \\
		\hline
Q2343-BM133	&	1.478	&	74.6$\pm1.2$	&	$\dots$	&	$\dots$	&	28.7$\pm0.4$	&	25.8$\pm2.0$	&	$\dots$	&	$\dots$	\\
Q2343-BM181	&	1.495	&	69.0$\pm5.6$	&	$\dots$	&	$\dots$	&	28.6$\pm0.7$	&	2.9$\pm0.6$	&	$\dots$	&	$\dots$	\\
Q2343-BX182	&	2.288	&	$\dots$	&	67.4$\pm3.2$	&	64.3$\pm11.8$	&	32.4$\pm0.5$	&	$\dots$	&	9.3$\pm1.6$	&	2.3$\pm0.5 $	\\
Q2343-BX236	&	2.434	&	$\dots$	&	81.6$\pm11.2$	&	$\dots$	&	29.8$\pm0.9$	&	$\dots$	&	1.8$\pm0.7$	&	$\dots$	\\
Q2343-BX336	&	2.545	&	$\dots$	&	66.8$\pm4.6$	&	$\dots$	&	26.7$\pm0.7$	&	$\dots$	&	4.4$\pm0.8$	&	$\dots$	\\
Q2343-BX341	&	2.576	&	$\dots$	&	76.6$\pm4.0$	&	$\dots$	&	28.1$\pm1.0$	&	$\dots$	&	10.9$\pm1.0$	&	$\dots$	\\
Q2343-BX389	&	2.172	&	$\dots$	&	64.7$\pm5.5$	&	58.0$\pm10.2$	&	31.6$\pm1.0$	&	$\dots$	&	4.6$\pm0.7$	&	1.3$\pm0.4$	\\
Q2343-BX390	&	2.232	&	$\dots$	&	74.1$\pm5.0$	&	76.5$\pm19.9$	&	30.5$\pm0.7$	&	$\dots$	&	4.8$\pm0.6$	&	1.3$\pm0.3$	\\
		\hline
		\hline
		\multicolumn{7}{l}{}\\
		\multicolumn{7}{l}{Flux units in 10$^{-17}$ erg s$^{-1}$ cm$^{-2}$.}\\
		\multicolumn{7}{l}{$^a$ Gaussian fits  using the PYTHON routine mpfit.}\\
		\multicolumn{7}{l}{$^b$ Measurements  using the IRAF task splot.}\\
		\multicolumn{7}{l}{The full version of this table is available as supplementary material.}\\
	\end{tabular}
\end{table*}

%%%%%%%%%%%%%%%%%%% Table 3 %%%%%%%%%%%%%%%
\begin{table*}
	\small
	\centering
	\caption{Total sample used for the cosmological analysis.}
	\label{tab:redshift,sigma and fluxes}	
		\begin{tabular}{lcccc}
			\hline
Target	&	z	&	log$\sigma$	&	logF(H$\beta$)	& note$^a$	\\
		\hline
	&		&		&		&		\\
		\multicolumn{5}{c}{\textbf{KMOS Sample}}							\\
	&		&		&		&		\\
COSMOS8991	&	2.2203	&	$1.725\pm0.023$	&	$-16.456\pm0.251$	&	1	\\
COSMOS11212	&	2.1984	&	$1.655\pm0.020$	&	$-16.497\pm0.150$	&	1	\\
	&		&		&		&		\\
		\multicolumn{5}{c}{\textbf{MOSFIRE Sample}}							\\
	&		&		&		&		\\
COSMOS-12807	&	1.5820	&	$1.726\pm0.039$	&	$-16.869\pm0.086$	&	2	\\
COSMOS-13848	&	1.4433	&	$1.447\pm0.085$	&	$-17.007\pm0.104$	&	2	\\
	&		&		&		&		\\
		\multicolumn{5}{c}{\textbf{XShooter Sample}}							\\
	&		&		&		&		\\
HoyosD2-1	&	0.8509	&	$1.703\pm0.045$	&	$-15.587\pm0.183$	&	3	\\
HoyosD2-5	&	0.6364	&	$1.616\pm0.010$	&	$-15.577\pm0.183$	&	3	\\
	&		&		&		&		\\
		\multicolumn{5}{c}{\textbf{Literature Sample}}							\\
	&		&		&		&		\\
COSMOS-17839	&	1.4120	&	$1.675\pm0.082$	&	$-16.714\pm0.428$	&	4	\\
GOODS-S-43928	&	1.4720	&	$1.565\pm0.098$	&	$-16.576\pm0.125$	&	4	\\
	&		&		&		&		\\
		\multicolumn{5}{c}{\textbf{Local Sample}}							\\
	&		&		&		&		\\
J001647$-$104742	&	0.0220	&	$1.377\pm0.039$	&	$-12.962\pm0.060$	&	5	\\
J002339$-$094848	&	0.0519	&	$1.463\pm0.036$	&	$-13.325\pm0.060$	&	5	\\
		\hline
			\multicolumn{5}{l}{Typical redshift uncertainty $\sim 10^{-4}$.}\\
			\multicolumn{5}{l}{Flux in erg s$^{-1}$ cm$^{-2}$ and velocity dispersion in km/s.}\\
			\multicolumn{5}{l}{$^a$ the flag corresponds to 1: KMOS, 2: MOSFIRE, 3: XShooter, 4: taken from}\\
			\multicolumn{5}{l}{the literature \citep{Erb2006a,Erb2006b, Masters2014, Maseda2014}, and}\\
			\multicolumn{5}{l}{5: local HIIG \citep{Chavez2014}.}\\
			\multicolumn{5}{l}{The full version of this table is available as supplementary material.}\\		
		\end{tabular}
\end{table*}

\end{document}